\documentclass[11pt,a4paper]{article}

\usepackage{hhline}
\usepackage[table]{xcolor}
\definecolor{maroon}{cmyk}{0,0.87,0.68,0.32}
\definecolor{Gray}{gray}{0.95}

\usepackage{graphicx}
\usepackage[english]{babel}

\usepackage[T1]{fontenc}
\usepackage{listings}

\usepackage{amsthm}
\usepackage{amsmath}
\usepackage{enumerate}
\usepackage{float}
\usepackage{graphicx}
\usepackage{subfigure}
\usepackage{setspace}
\usepackage{amssymb}
\usepackage{esint}
\usepackage{dcolumn}
\usepackage{booktabs}
\usepackage{eurosym}
\usepackage{longtable}
\usepackage{dcolumn}
\usepackage{hyperref}
\usepackage{fancyhdr}
\usepackage{multirow}
\usepackage{caption}
\usepackage{wrapfig}
\usepackage{rotating}
\usepackage{wrapfig}
\usepackage{afterpage}
\usepackage[utf8]{inputenc}
\usepackage{natbib}
\usepackage{bbm, dsfont}
\usepackage{subfigure}
\usepackage{booktabs}

\newcommand{\boldchi}{\mathcal{X}}
\newcommand{\boldx}{\mathcal{X}}
\newcommand{\X}{\mathcal{X}}
\newcommand{\bolddelta}{\boldsymbol{\delta}}

\newcommand{\boldy}{\mathcal{Y}}
\newcommand{\boldzeta}{R}

\newcommand{\vecm}[1]{{\overrightarrow {#1}}}

\newcommand{\Hsp}{\mathcal{H}}
\newcommand{\s}{s}
\newcommand{\h}{h}
\newcommand{\argmin}[1]{\underset{#1}{\operatorname{argmin}}\,\,}

\begin{document}

\title{\bf Kriging Riemannian Data via Random Domain Decompositions}

\author{Alessandra Menafoglio$^1$, Davide Pigoli$^2$\footnote{Address for correspondence: Davide Pigoli,
Department of Mathematics, King's College London, Strand, London WC2R 2LS, United Kingdom. Email:
davide.pigoli@kcl.ac.uk}\hspace{1mm} and Piercesare Secchi$^{1,3}$ \\\\
    \small $^1$MOX, Department of Mathematics, Politecnico di Milano, Milano, Italy\\
    \small $^2$Department of Mathematics, King's College London, London, United Kingdom\\
    \small $^3$Center for Analysis, Decisions and Society, Human Technopole, Milano, Italy\\
}

\date{}

\maketitle

%
%
\begin{abstract}
Data taking value on a Riemannian manifold and observed over a complex spatial domain are becoming more frequent in applications, e.g. in environmental sciences and in geoscience. The analysis of these data needs to rely on local models to account for the non stationarity of the generating random process, the non linearity of the manifold and the complex topology of the domain. In this paper, we propose to use a random domain decomposition approach to estimate an ensemble of local models and then to aggregate the predictions of the local models through Fr\'{e}chet averaging. The algorithm is introduced in complete generality and is valid for data belonging to any smooth Riemannian manifold but it is then described in details for the case of the manifold of positive definite matrices, the hypersphere and the Cholesky manifold. The predictive performance of the method are explored via simulation studies for covariance matrices and correlation matrices, where the Cholesky manifold geometry is used. Finally, the method is illustrated on an environmental dataset observed over the Chesapeake Bay (USA).
\end{abstract}


\section{Introduction} 

In many problems of applied interest, data can be better understood if embedded in a non Euclidean space. Statisticians have been aware for a long time of the particular challenges these data present, indeed since the first works on spherical data (\citet{fisher1953dispersion}) and shapes (\citet{kendall1977diffusion}), motivated by practical problems in astronomy, anthropology and geology. It is now recognised that considerable care needs to be taken to understand what are the key properties of the statistical units -- i.e., the atoms of the analysis -- which should be accounted for in a sensible statistical analysis. This viewpoint indeed provides the foundation of Object Oriented
Data Analysis (OODA, \citet{marron2014overview}), a system of ideas and methods for the statistical analysis of complex data. In this context, a considerable effort has been made to develop comprehensive frameworks for the analysis of general types of data, taking into account the geometry of the space (see,  e.g., \cite{patrangenaru2015nonparametric}).

Although the literature on OODA is nowadays well-developed, a question that is yet to be fully addressed is how to measure, and incorporate  in the analysis, the stochastic dependence between observations. This point is particularly challenging when data are spatially distributed, i.e., in the context of Object Oriented Spatial Statistics (O2S2, \citet{MenafoglioSecchi2017}), and possibly generated by a non-stationary stochastic process. For instance, when data are manifold-valued objects, existing approaches to spatial prediction have been mainly based on a tangent space approximation (i.e., linearisation) of the data, that enables one to rely on linear techniques in O2S2 while accounting for the mildly non Euclidean structure of Riemannian manifolds (see \cite{PigoliEtAl2016}). These approaches have clear limitations whenever the variability of the observations is particularly large and complex, which may be well the case in the presence of large spatial domains, or complex non-stationarities.

In parallel to this literature stream, recent research in spatial statistics has started to address the problem of analysing spatial data when their domain of observation has a complex topology, e.g., in the presence of irregular boundaries, holes and barriers (e.g., \citet{Ramsay2002,WoodEtAl2008,SangalliEtAl2013,BernardiEtAl2015, MenafoglioEtAl2018} and references therein). Even though the challenges arising from this latter problem are indeed different from those related with the analysis of manifold data, the two problems share the need to \emph{localize} the models for the data analysis.

In this work, we jointly tackle these challenges through the key role of random domain decompositions (RDDs), which are here used to estimate an ensemble of \emph{local} models. The proposed approach extends to Riemannian data the ideas of \citet{MenafoglioEtAl2018}, which were developed in the context of non-stationary spatial fields of Hilbert data. The computational methodology we propose is based on a bagging strategy which consists of two steps: (i) repeatedly and randomly partition the domain of observation in disjoint subdomains, where to estimate \emph{local} geostatistical models and perform prediction (Kriging) and (ii) aggregate the results of these repetitions to provide a final result. Unlike \citet{MenafoglioEtAl2018}, the RDD is here used to support the construction and estimate of a local linearization of the observations, which is then employed for modeling and prediction purposes.

Although the idea of using domain decompositions has similarities with existing machine learning techniques (e.g., \citet{GramacyLee2008},\citet{RasmussenGrahramani}), a bagging approach is used in this work to control the uncertainty and possible spatial discontinuities introduced by the random partitioning of the spatial domain.

The proposed method is of general validity, being well-suited for any kind of Riemannian data. We here describe its implementation in detail for two specific examples: positive definite matrices (covariance matrices) and correlation matrices (and, more generally, hyperspheres). We use these two examples as test cases to assess the performances of the method in a simulation setting. We finally illustrate the application of the method to a dataset of environmental interest, consisting of covariance matrices estimated within the Chesapeake estuarine system, which is a domain with complex topology.

The remaining part of the paper is structured as follows. Section \ref{sec:recall-RDD} recalls the RDD method for Hilbert data of \citet{MenafoglioEtAl2018}, whereas Section \ref{sec:recall-man} provides the modeling framework for the development of our method. Section \ref{sec:RDD-man} describes the RDD method for kriging manifold data (RDD-MK), which is described in detail for the cases of covariance and correlation matrices in Section \ref{sec:geometries}. Section \ref{sec:simu} illustrates the simulation study and Section \ref{sec:case-study} the case study. Section \ref{sec:concl} concludes the work. 

\section{RDD for kriging in Hilbert spaces}\label{sec:recall-RDD}

We here recall the computational methodology proposed by \citet{MenafoglioEtAl2018} to perform kriging of Hilbert data, distributed over domains with complex topology (irregular boundaries, holes, barriers).

Let $(\Omega, \mathcal{F}, \mathbb{P})$ be a probability space, and call $D\subset\mathbb{R}^d$ (usually $d=2,3$) the spatial domain of interest. In the same framework considered by \citet{MenafoglioEtAl2018}, denote by $\{\boldy_\s, \s\in D\}$ a random field defined on $(\Omega, \mathcal{F}, \mathbb{P})$ and with values in a separable Hilbert space $\Hsp$. Given $n$ locations $\s_1,...,\s_n$ in $D$, denote by $\boldy_{\s_1},...,\boldy_{\s_n}$ the observations of the field at these locations, which are (random) elements of $\Hsp$.

Whenever the domain $D$ is simple and the random field $\{\boldy_\s, \s\in D\}$ can be assumed to be stationary, a mathematical framework for kriging can be established as in \citep{MenafoglioEtAl2013,MenafoglioSecchi2017}, where the well-known concepts of covariogram, variogram and kriging predictor \citep[e.g.][]{Cressie1993} were extended to the Hilbert space setting. The latter framework can also be considered in the presence of mild non-stationarities, i.e., when the non-stationarity can be modeled by decoupling each element $\boldy_\s$, $\s$ in $D$, as the sum of a mean term $m_\s$ -- described by a linear model -- and a stationary residual $\bolddelta_\s = \boldy_\s - m_\s$.

In case of strong non-stationarity of the random field or of a domain $D$ with a complex topology (irregularly shaped, with holes or barriers), the framework of \citet{MenafoglioEtAl2013} cannot be used, as it is not able to capture (i) a possible non-stationarity of the residual random field and (ii) a non-Euclidean metric on the spatial domain.
However, the degree of non-stationarity of the field, as well as the degree of complexity of the domain, may depend on the spatial scale of observation. Indeed, even though the field may appear non-stationary at a global spatial scale, stationarity may be a viable assumption at a local scale (i.e., local stationarity). Similarly, a spatial domain which is complex (non-Euclidean) at a global scale, might be locally approximated through a simple (Euclidean) domain. The random domain decomposition (RDD) approach introduced by \citet{MenafoglioEtAl2018} brings the latter foundational idea into play by implementing an operational strategy, named Random Domain Decomposition for Object Oriented Kriging (RDD-OOK).

RDD-OOK is grounded on a \emph{divide-et-impera} approach, and follows a bagging strategy \citep{Breiman1996}. In the \emph{bootstrap} stage, the domain is iteratively randomly partitioned in tiles, where local stationary analyses are performed. For a given location $\s_0$, this stage results in a set of $B$ predictions, where $B$ stands for the number of bootstrap replicates. Each of these $B$ predictions is conditioned on the realization of the RDD, and is based on the stationary model estimated within the tile at that iteration. Finally the $B$ predictions resulting from the bootstrap stage are aggregated in a final prediction (a.k.a. \emph{aggregation} step), typically by taking a (weighted) average of the bootstrap predictions. For further details, we refer the reader to \citet{MenafoglioEtAl2018}. In the following sections, we extend this framework to Riemannian data, and detail the modeling choices adopted in this work.

\section{Statistical analysis of Riemannian data}\label{sec:recall-man}
Data belonging to Riemannian manifolds appear in many applications and they have long been studied in statistics. Examples include but are not limited to directional data (\cite{fisher1993statistical}, \cite{mardia2014statistics}), shapes (\cite{goodall1991procrustes}, \cite{dryden1998statistical}) and positive definite matrices (\cite{Dryden2009}, \cite{Yuan2012}). We are going to discuss here only the essential ingredients of the Riemannian geometry that need to be taken into account when developing the kriging prediction. For a complete description of the state of the art of statistics on manifolds we refer the interested reader to \citet{patrangenaru2015nonparametric}.

\subsection{Riemannian manifolds: preliminaries and definitions}
A manifold is a metric space which is locally homeomorphic to an Euclidean space (see, e.g., \citet{patrangenaru2015nonparametric}). Given a point $\Psi$ in $\mathcal{M}$, we denote by $T_\Psi\mathcal{M}$ the tangent space at $\Psi$ to $\mathcal{M}$.  For a manifold to be Riemannian, it means that, for every $\Psi \in \mathcal{M}$, it is possible to define a bilinear form $\kappa(\Psi): T_\Psi\mathcal{M} \times T_\Psi\mathcal{M} \rightarrow \mathbb{R}$, called Riemannian structure, that is symmetric, positive definite and depends smoothly on $\Psi$. Then, $T_\Psi\mathcal{M}$ is a Hilbert space when equipped with the inner product $\langle\cdot,\cdot\rangle_{\Psi}=\kappa(\Psi)(\cdot,\cdot)$. If the manifold $\mathcal{M}$ is connected (i.e., it cannot be expressed as disjoint union of two non-empty open subset), the Riemannian structure (and hence the inner product on the tangent space) induces a metric on the manifold $\mathcal{M}$, in the following way. Let $\mathcal{P}(\Psi_1,\Psi_2)$ be the set of piecewise differentiable curves $c: [0,1] \rightarrow \mathcal{M}$  such that $c(0)=\Psi_1$ and $c(1)=\Psi_2$. The (kinetic) energy functional $E: \mathcal{P}(\Psi_1,\Psi_2) \rightarrow \mathbb{R}$, is defined by
$$
E(c)=\frac{1}{2}\int_{0}^1 \kappa(c(t))\Big(\frac{\mathrm{d}c(t)}{\mathrm{d}t},\frac{\mathrm{d}c(t)}{\mathrm{d}t}\Big)\mathrm{d}t= \frac{1}{2}\int_{0}^1 \left\lVert\frac{\mathrm{d}c(t)}{\mathrm{d}t}\right\rVert^2_{c(t)}\mathrm{d}t.
$$
Here the norm $||a||_{c(t)}=\sqrt{\langle a, a \rangle_{c(t)}}$ is the one induced by the inner product on $T_{c(t)}\mathcal{M}$. A \emph{geodesic} from $\Psi_1$ to $\Psi_2$ is the path $g$ in $\mathcal{P}(\Psi_1,\Psi_2)$ that minimises the energy functional and the \emph{geodesic distance} $d_R$ between $\Psi_1$ and $\Psi_2$ is the length of this path, i.e.
$$
d_R(\Psi_1,\Psi_2)= \int_0^1 \left\lVert\frac{\mathrm{d}g(t)}{\mathrm{d}t}\right\rVert_{g(t)}\mathrm{d}t.
$$

\begin{figure}
  \centering
  \includegraphics[height=4cm]{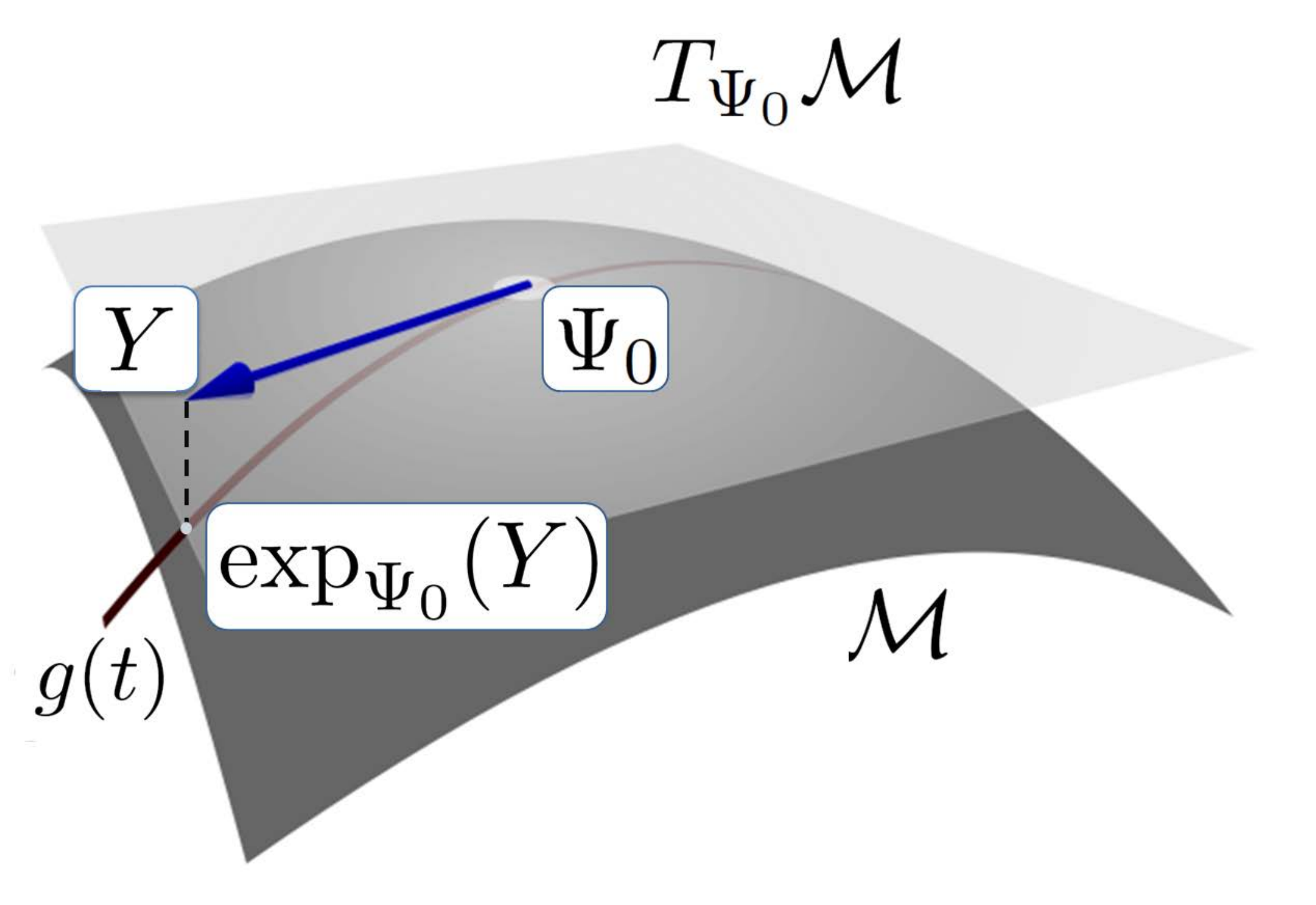}\hspace*{1cm}
  \includegraphics[height=4cm]{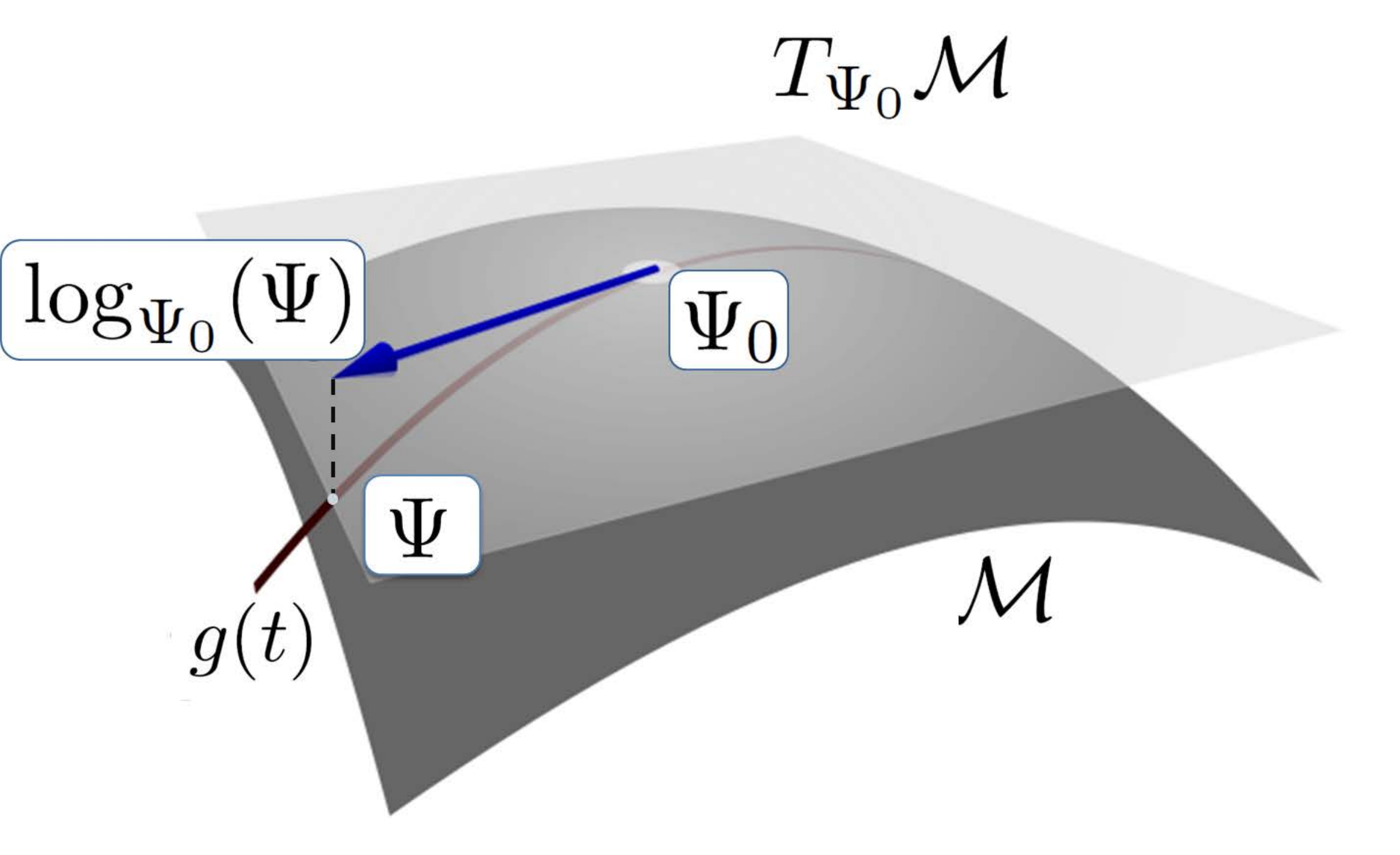}\\
  \caption{Representation of exponential and logarithmic maps.}\label{fig:exp-log-maps}
\end{figure}

For the scope of this work, we also assume that the manifold $\mathcal{M}$ is geodesically complete, i.e. for any pair of points $\Psi_1,\Psi_2 \in \mathcal{M}$ it exists a geodesic joining $\Psi_1$ and $\Psi_2$. Then, from the theory of second order differential equations, it follows that for every pair $(\Psi_0,Y)$ in $\mathcal{M}\, \times\, T_{\Psi_0}\mathcal{M}$, there exists a unique geodesic curve $g(t)$ such that $g(0) = \Psi_0$ and $\frac{\mathrm{d}g}{\mathrm{d}t}(0) = Y$. In this context, one can define exponential and logarithmic maps, that, based on $g$, associate objects in $T_{\Psi_0}\mathcal{M}$ with objects in $\mathcal{M}$ and viceversa. The exponential map is defined as a smooth function from $T_{\Psi_0}\mathcal{M}$ to $\mathcal{M}$, that maps a tangent vector $Y \in T_{\Psi_0}\mathcal{M}$, to the point at $t = 1$ of the geodesic starting in $\Psi_0$ with direction $Y$, i.e.,
$$
\exp_{\Psi_0}(Y)=g(1).
$$
The exponential map is invertible in a neighbourhood of $0$; we denote its inverse as the logarithmic map
$\log_{\Psi_0}(\Psi)$ which returns the tangent point $Y$ in $t=0$ associated with the geodesic such that $g(0)=\Psi_0$ and $g(1)=\Psi$. Figure \ref{fig:exp-log-maps} shows a visualisation of these latter notions.
Locally, the geodesic passing through $\Psi_0$ and $\Psi$ is represented by a line going through the origin, i.e. $g(t)=t \log_{\Psi_0}(\Psi)$ and the distance with respect to $\Psi_0$ is preserved as $d_R(\Psi_0,\Psi)=\sqrt{\langle \log_{\Psi_0}(\Psi), \log_{\Psi_0}(\Psi) \rangle_{\Psi_0}}$.

We can then look at the largest domain for which $\exp_{\Psi_0}(\cdot)$ is a diffeomorphism. Let $t_0$ be the largest $t$ for which the geodesic $g$ is minimising the path length between $\Psi_0$ and $g(t)$, then $g(t_0)$ is called a cut point and the set of all cut points for all the geodesics starting from $\Psi_0$ is called the cut locus $C(\Psi_0)$. The set of tangent vectors $\mathcal{TC}(\Psi_0)$ such that $C(\Psi_0)=\exp_{\Psi_0} (\mathcal{TC}(\Psi_0))$ is called tangent cut locus and the maximal definition domain for the exponential map is the domain that contains $0$ and is delimited by the tangent cut locus.

We refer the reader to \citep{Lee2012} for further details on the definitions and on the properties of Riemannian manifolds; the Riemannian manifolds of covariance and correlation matrices are further detailed in Subsections \ref{subsubsec:cov} and \ref{subsubsec:cor}, respectively.

\subsection{Random elements in a Riemannian manifold}\label{subsec:stat-manifold}
Let us now consider a random variable $\boldx$ taking value on the Riemannian manifold $\mathcal{M}$, i.e. a measurable map on a probability space $(\Omega, \mathcal{F}, \mathbb{P})$ into $(\mathcal{M},\mathcal{B})$, where $\mathcal{B}$ is the Borel sigma-algebra generated by the open subsets of $\mathcal{M}$. The probability distribution $P_{\boldx}$ on $\mathcal{B}$ associated to $\boldx$ can be defined as $P_{\boldx}(B)=\mathbb{P}(\boldx^{-1} (B))$, for $B$ in $\mathcal{B}$. We can then define the moments of this random variable via a \emph{Fr\'{e}chet function} based on the Riemannian distance $d_R$. For example,
\begin{equation}\label{eq:frechet-function}
\centering
\mathcal{F}(\Psi)=\int d_R^2(\Psi,x)P_{\boldx}(\mathrm{d}x),
\end{equation}
for each $\Psi \in \mathcal{M}$. The minimiser(s) of $\mathcal{F}$ is called Fr\'{e}chet mean (set) or the \emph{intrinsic mean} (set) of $\boldx$.  It was first defined by \citet{frechet1948elements} as the set of global minima of the variance in a general metric space and later used to summarise data points on Riemannian manifolds by \citet{karcher1977riemannian}. If the $\mathcal{F}(\Psi)$ is finite for some $\Psi$, then the intrinsic mean set is non empty. The conditions for the intrinsic mean to be unique are more restrictive. The minimum is unique for simply connected Riemannian manifolds of non-positive curvature (see, e.g., \cite{pennec2006intrinsic}, for more details). This is the case for example of the manifold of positive definite matrices described in Section \ref{subsubsec:cov}. There are also alternative conditions that guarantee the intrinsic mean to be unique if the support of the random variable $\boldx$ is localised, see \citet{karcher1977riemannian} and \citet{kendall1990probability} for the technical details. For the purpose of the random domain decomposition algorithm, this would require to define tiles that are small enough so that the variation of the data within the tile is also small.

If the Fr\'{e}chet function \eqref{eq:frechet-function} of $\boldx$ is finite, its local minima (and therefore also its global minimum, the Fr\'{e}chet mean) can be characterised in the following way (\citet{pennec2006intrinsic}). Let $\mathcal{A}$ be the set of points where the cut locus has non zero probability measure, then any local minimum $\overline{\Psi}$  of the Fr\'{e}chet function is such that either $\overline{\Psi}\in \mathcal{A} $ or $E[\log_{\overline{\Psi}}(\boldx)]=0$ if $\overline{\Psi} \in \mathcal{A}^C$.

Given a sample $\boldx_1,\ldots,\boldx_n$ from the random variable $\boldx$, it is possible to define the empirical equivalent of the Fr\'{e}chet mean as the sample Fr\'{e}chet mean
$$
\widehat{\Psi}=\argmin{\Psi} \sum_{i=1}^n d_R^2(\boldx_i,\Psi),
$$
also called sample intrinsic mean. If the intrinsic mean exists and is unique, the sample intrinsic mean is a consistent estimator. This is true even when the intrinsic mean is not unique if we accept a broader definition of consistency,  see \citet{bhattacharya2003large} for more details and an extensive discussion of the properties of this estimator. In particular, the characterization of local minima for the Fr\'{e}chet function described above still holds when $P_{\boldx}$ is the empirical probability distribution generated by the observations. Now the set $\mathcal{A}$ is given by the union of the cut loci for all the observed data points $\boldx_1,\ldots,\boldx_n$ and, if the local minimum $\overline{\Psi}$ does not belong to $\mathcal{A}$, then $\sum_{i=1}^n \log_{\overline{\Psi}}(\boldx_i)=0$.

\subsection{A tangent space model for spatial manifold data}
In the broad context of Subsection \ref{subsec:stat-manifold}, a geostatistical framework for kriging Riemannian data can be based on a tangent-space approximation.

We now denote by $\boldx_{\s_1},..., \boldx_{s_n}$ the observations available at locations $\s_1,...,\s_n$ in $D$, which are here interpreted as a partial observation of a random field $\{\boldx_{\s}, \s\in D\}$, valued in a Riemannian manifold $\mathcal{M}$. Given a location $s_0$ in $D$, we consider the problem of performing spatial prediction of the unobserved element $\boldx_{\s_0}$.
A possible model for the random element $\boldx_\s$ is
\begin{equation}\label{eq:model-pigoli}
\boldchi_\s(\Psi_\s) = \exp_{\Psi_\s}\left\{m_\s+\bolddelta_\s\right\}, \quad \s\in D,
\end{equation}
where $\Psi_\s$ is a point in $\mathcal{M}$, $m_\s$ is the mean in $T_{\Psi_s}\mathcal M$ and $\bolddelta_\s$ a zero-mean process in $T_{\Psi_\s}\mathcal{M}$.

Specializing model \eqref{eq:model-pigoli}, one may derive a variety of statistical methods for manifold data.  For instance, the geodesic regression model in \citet{fletcher2013geodesic} falls within this framework for the case of independent errors and a specific parametric model for $\Psi_\s$. For the case we are considering in this work, it is of particular relevance the geostatistical model described in \citet{PigoliEtAl2016}, which is a special case of model \eqref{eq:model-pigoli}, with the further assumptions that (i) the tangent point $\Psi_\s$ is spatially constant (i.e., $\Psi_\s=\Psi$ for $\s\in D$), (ii) the mean $m_\s$ is described by a linear model in $T_{\Psi\s} \mathcal{M}$, $m_\s = A\left(\vecm f({\s}),\vecm{a}\right)$,  with scalar regressors $\vecm f({\s}) = (f_0(\s),...,f_L(s))^T$ and coefficients $\vecm{a} = (a_0,...,a_L)$, $a_l\in\mathcal{M}$, $l=0,...,L$, and (iii) the stochastic process of the residuals $\{\bolddelta_\s,\s\in D\}$ is second-order stationary in the sense of \citep{MenafoglioEtAl2013}, with trace-variogram
\begin{equation}\label{eq:tr-var}
2{\gamma}(\s_i-\s_j) = \mathbb{E}[\|\bolddelta_{\s_i}-\bolddelta_{\s_j}\|_{\Psi}^2], \quad\textrm{for }\s_i,\s_j\in D.
\end{equation}
We recall that, in the Hilbert setting, the trace-variogram \eqref{eq:tr-var} describes the global second order properties of the field; as such, it plays the same key role as its classical counterpart \citep[e.g.,][]{Cressie1993}.
Although model \eqref{eq:model-pigoli} is quite general, it should be noted that its specialization to the model of \citep{PigoliEtAl2016} allows to accurately represent the geometry of the data only locally on the manifold (i.e., in a neighborhood of $\Psi$). Thus, it is apt to describe fields characterized by a limited variability on the manifold. Further, as in the case of Hilbert data \citep{MenafoglioEtAl2013}, the model of \citep{PigoliEtAl2016} may capture the possible non-stationarity of the field only by modeling the drift through the term $A(\vecm f({\s}),\vecm{a})$, the residuals being assumed stationary.

We here aim to develop a methodology to allow for (i) a high degree of variability of the data on the manifold; (ii) a strong non-stationarity of the field; (iii) a possible complex domain topology which hinders the use of state-of-the-art geostatistical approaches for manifold data. To cope with all these issues jointly, we here consider model \eqref{eq:model-pigoli} to hold only in local spatial neighborhoods.
In practice, following the strategy originally proposed by \citet{MenafoglioEtAl2018}, our approach is based on the idea of using random decompositions of the domain to define the latter system of neighborhoods, estimate the local model conditionally on the RDD, and perform the kriging predictions accordingly. The next section provides a complete description of the proposed procedure.

\section{RDD for manifold data}\label{sec:RDD-man}

The methodology we propose follows a bagging strategy \citep{Breiman1996}. In the bootstrap step, the domain is randomly decomposed in subdomains where local tangent-space models are formulated and estimated. These models are used to provide a prediction conditional on the RDD. In the aggregation stage, the results of the bootstrap iterations are aggregated to provide a final prediction result. In the following paragraphs we detail the methodological proposal as well as the modeling choices of this work.

\subsection{Bootstrap stage}
\paragraph*{Local tangent-space approximations}
We call $\mathcal{P}$ an RDD of the domain $D$ in $K$ tiles, and denote by $P$ a realization of $\mathcal{P}$. For simplicity, we here follow the proposal of \citet{MenafoglioEtAl2018}, and consider as $\mathcal{P}$ a random Voronoi tessellation induced by a set of $K$ centers $\Phi_K=\{c_1,...,c_K\}$ randomly drawn among the observation sites $\{\s_1,...,\s_n\}$:
\begin{equation}\label{eq:voronoi}
V(c_k|\Phi_K) = \{\s\in D: d(\s,c_k) \leq d(s, c_j), \textrm{ for all }c_j\in\Phi_K, \; j\neq k\},\end{equation}
with $d$ a distance on $D$. Note that $d$ should be a distance properly representing proximity in $D$; in case of domains with a complex topology, $d$ could thus be different from the Euclidean distance. We remark that other systems of partitions can be considered as well. For instance, if prior knowledge is available on possible discontinuities in the spatial distribution of the field, this should be formalized by building $\mathcal P$ accordingly.

Given $\mathcal P  =P$, we call $D_1,...,D_K$ the $K$ tiles identified by the partition $P$. In each tile $D_k$, $k=1,...,K$, we model the field as in \citet{PigoliEtAl2016}. Indeed, for $\s \in D_k$ we consider model \eqref{eq:model-pigoli}, and set the tangent point $\Psi_\s$ to be constant, i.e., $\Psi_\s = \Psi_k$, for $k=1,...,K$.
In view of our application, hereafter we assume that the mean is constant within the tile $D_k$, i.e., $m_\s = m_k$ for $\s\in D_k$. Nonetheless, if covariates $\vecm f({\s})$ are available, these can be incorporated by further modeling the drift term, e.g., in a linear model $m_\s = A\left(\vecm f({\s}),\vecm{a}\right)$ as in \citep{PigoliEtAl2016}. We denote by $2{\gamma}(\s_i,\s_j;k)$ the trace-variogram of the residuals within the tile. By virtue of the stationary assumption within the tile, the latter trace-variogram reads
\begin{equation}\label{eq:tr-var-rdd}
2{\gamma}(\s_i-\s_j;k) = \mathbb{E}[\|\log_{\Psi_k}(\boldchi_{\s_i})-\log_{\Psi_k}(\boldchi_{\s_j})\|_{{\Psi}_k}^2], \quad \s_i,\s_j \in D_k.
\end{equation}
Note that the viability of the stationarity assumption may depend on the size of the tiles; here, the distribution $\mathcal{P}$ of the RDD is relevant to control the degree of \emph{locality} of the model.

We also remark that definition \eqref{eq:tr-var-rdd} strongly depends on the metric used in the tangent-space, which in turn depends on $\Psi_k$. In fact, model \eqref{eq:model-pigoli} entails the use of a different Hilbert metric for the observations of the field relative to locations in $D_k$, $k=1,...,K$, each metric being chosen as to best represent the geometrical structure of the elements in the tile. This flexibility clearly allows for an enhanced characterization of the geometry of the Riemannian manifold with respect to previous works, which may be particularly relevant for data characterized by a large variability in $\mathcal{M}$.

\paragraph*{Model Estimation Conditional on the RDD}
Given a realization of the RDD, we here propose to estimate model \eqref{eq:model-pigoli} in two steps: (i) estimate the local tangent point $\widehat{\Psi}_k$, $k=1,...,K$, (ii) given $\widehat{\Psi}_k$, estimate a model for the spatial dependence in $\Hsp_{\widehat{\Psi}_k}$, $k=1,...,K$.

For $k=1,...,K$, we propose to estimate $\Psi_k$ via the intrinsic mean of the observations within the tile $D_k$, i.e., as
\begin{equation}\label{eq:Psi_hat}
\widehat{\Psi}_k = \argmin{\Psi \in \mathcal{M}} \sum_{\s_i \in D_k} d^2_R(\Psi,\boldchi_{\s_i}),
\end{equation}
where $d_R(\cdot,\cdot)$ denotes the Riemannian distance in $\mathcal{M}$, i.e. estimator \eqref{eq:Psi_hat} is the sample Fr\'{e}chet mean (or the sample intrinsic mean) of the data within the tile. In general, no explicit expression for the sample intrinsic mean is available, but the latter can be found by implicit optimization routines. 

Given $\widehat{\Psi}_k$, we estimate the model for the spatial dependence in the Hilbert space $\Hsp_{\widehat{\Psi}_k}$ by extending the ideas of \citet{MenafoglioEtAl2018}. By virtue of stationarity in $D_k$, we estimate the trace-semivariogram $\gamma(\cdot;k)$ via a geographically weighted estimator as
\begin{equation}\label{eq:tr-var-est}
\widehat{\gamma}(\h;k) = \frac{\sum_{N(\h)} \mathcal{K}(c_k,\s_i)\mathcal{K}(c_k,\s_j) \|\log_{\widehat{\Psi}_k} (\boldchi_{s_i})-\log_{\widehat{\Psi}_k} (\boldchi_{s_j})\|_{\widehat{\Psi}_k}^2}{2\sum_{N(\h)}
\mathcal{K}(c_k,\s_i)\mathcal{K}(c_k,\s_j)},
\end{equation}
where $N(\h)=\{(\s_i,\s_j)\in D\times D: \h-\Delta\h \leq \s_i-\s_j \leq \h + \Delta\h\}$, and $\mathcal{K}$ is a kernel function. Although \citet{MenafoglioEtAl2018} use a Gaussian kernel (i.e., $\mathcal{K}_\epsilon(\s_1,\s_2) = \exp\{-d^2(\s_1,\s_2)/(2\epsilon^2)\}$), different choices are indeed possible. One may also opt for a trivial kernel which associates weight 1 to the observations within the tile, and 0 otherwise; in this case, the estimate $\gamma(\cdot;k)$ would only be based on the pairs within $D_k$. In general, estimator \eqref{eq:tr-var-est} will down-weight the contribution of pairs whose locations are far from the center of the tile. This is consistent with the local nature of the tangent-space approximation as well as of the stationarity assumption. Nonetheless, estimator \eqref{eq:tr-var-est} allows to borrow strength from neighboring tiles, thus increasing its robustness in tiles of small size.
Whatever the choice of the kernel, to guarantee the validity of the estimated variogram, a parametric valid model (e.g., spherical, exponential, Mat\'{e}rn)  $\gamma(\cdot;{\theta}, k)$ should be fitted to $\widehat{\gamma}(\cdot;k)$ (e.g., by least squares). The final estimate of the spatial dependence within $D_k$, ${\gamma}(\cdot; \hat{\theta}, k)$ can be then used for the purpose of prediction of the unobserved element $\boldchi_{\s_0}$.

\paragraph*{Kriging Conditional on the RDD}
Given the estimated variogram model, we build the optimal predictor for $\boldchi_{\s_0}$, $\s_0 \in D_k$, as the image in $\mathcal{M}$ of the kriging predictor in $\Hsp_{\widehat\Psi_k}$ under model \eqref{eq:model-pigoli} restricted to $D_k$. This is defined as
\begin{equation}
\label{eq:kriging}
\boldchi^{*}_{\s_0} = \exp_{\widehat{\Psi}_k}\left(\sum_{i=1}^{n}\lambda^{*}_i\cdot\log_{\hat\Psi_k}(\boldchi_{\s_i}) \mathbbm{1}\{\s_i \in D_k\}\right),
\end{equation}
where $\mathbbm{1}$ is the indicator function, and the weights $\lambda^{*}_1, \dots, \lambda^{*}_n \in \mathbb{R}$ minimize
\begin{gather}\label{eq:kriging-min}
\nonumber\mathbb{E} \left[\left\|\log_{\hat\Psi_k}(\boldchi_{\s_0}) - \sum_{i=1}^{n}\lambda_i\cdot \log_{\hat\Psi_k}(\boldchi_{\s_i})\mathbbm{1}\{\s_i \in D_k\}\right\|^2_{\hat\Psi_k}\right] \\
\textrm{ subject to }\\
\nonumber\mathbb{E}\left[\sum_{i=1}^{n}\lambda_i\cdot \log_{\hat\Psi_k}(\boldchi_{\s_i})\mathbbm{1}\{\s_i \in D_k\}\right]=m_{k},
\end{gather}
over $\lambda_1, \dots, \lambda_n \in \mathbb{R}$. The solution of system \eqref{eq:kriging-min} exists and it is unique; it can be explicitly found by solving a linear system that only depends on the (estimated) trace-semivariogram ${\gamma}(\cdot; \hat{\theta}, k)$. For further details on kriging in Hilbert spaces we refer the reader to \citep{MenafoglioEtAl2013,menafoglio2016kriging,MenafoglioSecchi2017}.

\subsection{Aggregation stage}

The result of the bootstrap iterations is a set of kriging predictions $\{\boldchi^{*b}_{\s_0}\}_{b=1}^B$. Each of these predictions is obtained conditioned on the realization of the RDD, and it is relative to the model \eqref{eq:model-pigoli} estimated from that iteration. To obtain a final prediction at location $\s_0$ one needs to aggregate this ensemble of predictors, to obtain a final strong result out of the $B$ weak results of the bootstrap stage. For this purpose, we here propose to use the intrinsic mean of the ensemble, that is
\begin{equation}\label{eq:aggregation}
\boldchi^{*}_{\s_0} = \argmin{x \in \mathcal{M}} \sum_{b=1}^B d^2_R(x,\boldchi^{*b}_{\s_0}).
\end{equation}
Weighted versions of \eqref{eq:aggregation} can be considered to associate a different weight to each iteration; this latter approach is not pursued further here.

\medskip

The pseudo-code of the proposed computational method is displayed in Figure \ref{fig:pseudo-code}.
In Section \ref{sec:geometries} we shall detail two paradigmatic examples of manifold data, namely covariance and correlation matrices, and embed the general proposed framework in these geometries, that are particularly relevant for the applications.
\bigskip

\edef\myindent{\the\parindent}
\framebox[1\textwidth]{
\begin{minipage}{0.9\textwidth}\setlength{\parindent}{\myindent}

\vspace*{0.4cm}
\begin{center}
\textbf{\underline{Manifold Kriging via RDD}}
\end{center}
\small
\begin{description}
\item[\underline{$\mathtt{Initialization.}$}]\ \\
Set the parameters $1\leq K\leq n$, $B\geq 1$, a kernel $\mathcal{K}$ with its parameters, a valid variogram model, a metric $d$ for the spatial domain $D$ and the target location $\s_0$.
\item[\underline{$\mathtt{Bootstrap}$  $\mathtt{step.}$}] \ \\  $\mathtt{for}$ $b:=1$ $\mathtt{to}$ $B$ $\mathtt{do}$
\begin{description}
\item[$\mathtt{Step}$ $\mathtt{1}$.] Draw a realization of \(\mathcal{P}.\)\\
Randomly generate a set of nuclei $\Phi_K=\{\mathbf{c}_1,\dots,\mathbf{c}_K \}$  among the observed sites $\s_1,...,\s_n \in D$; define the Voronoi cells $\{V(\mathbf{c}_k|\Phi_K)\}_{k=1}^K$ by assigning each site $\s$ to the nearest nucleus $\mathbf{c}_k$, according to the metric $d$.
\item[$\mathtt{Step}$ $\mathtt{2}$.] For each Voronoi cell $D_k$: estimate the tangent point $\Psi_k$ as in \eqref{eq:Psi_hat}; estimate the semivariogram $\hat{\gamma}(\mathbf{h};k)$, by means of \eqref{eq:tr-var-est}; fit the parametric valid model to the empirical estimate and obtain ${\gamma}(\cdot; \hat{\theta}, k)$.
\item[$\mathtt{Step}$ $\mathtt{3}$.] For \(\s_0 \in D_k,\) obtain the kriging prediction $\X^{*b}_{\s_0}$, as in \eqref{eq:kriging}.
\end{description}
\texttt{end for.}
\item[\underline{$\mathtt{Aggregation}$  $\mathtt{step.}$}]\ \\
For \(\s_0 \in D,\) compute the final prediction (RDD-MK predictor) by aggregating the $B$ predictions as their intrinsic mean
\[
\X^{*}_{\s_0} = \argmin{\boldchi \in \mathcal{M}}\sum_{b=1}^B d^2_R(\boldchi, \X^{*b}_{\s_0}).\]
\end{description}
\end{minipage}
}\hfill

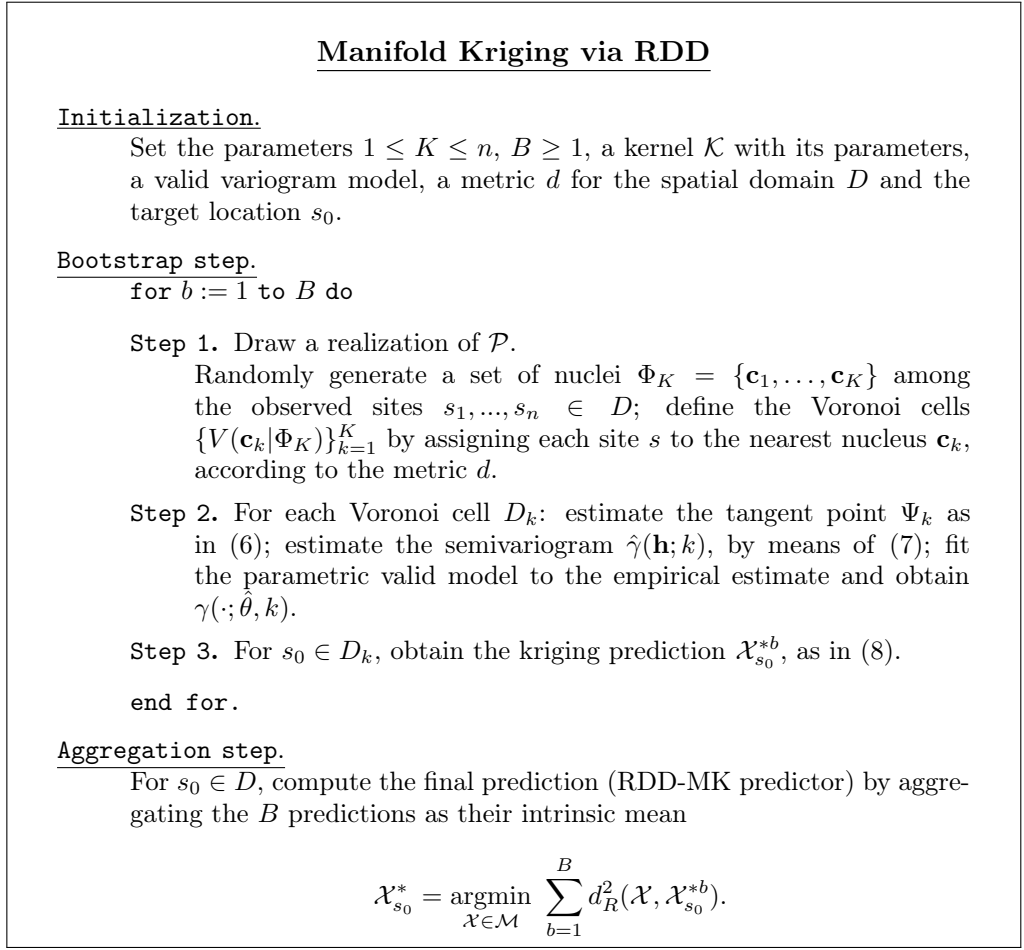
\captionof{figure}{Pseudocode scheme of the algorithm for Manifold Kriging through Random Domain Decomposition (RDD-MK).\label{fig:pseudo-code}}

\setlength{\parindent}{\myindent}

\section{Two paradigmatic examples}\label{sec:geometries}

\subsection{The space of covariance matrices}\label{subsubsec:cov}

We detail here our method in the special case where the Riemannian manifold $\mathcal{M}$ is the space $PD(p)$ of the $p \times p$ symmetric positive definite matrices. The space $Sym(p)$ of the $p\times p$ symmetric matrices is a Hilbert space that is usually equipped with the Frobenius inner product $\langle Y_1,Y_2\rangle_{Sym(p)}=\mathrm{trace}(Y_1^TY_2)$, which induces the well-known Frobenius norm $||Y||_{Sym(p)}=\sqrt{\mathrm{trace}(Y^TY)}=\sqrt{\sum_{i,j} Y_{i,j}^2}$. The additional constraint on the matrices to be positive definite makes instead the space $PD(p)$ a non Euclidean space, even if it can be embedded in $Sym(p)$.    
Different metrics have been proposed to measure differences between positive definite matrices \citep[][]{Dryden2009} but a natural choice is a metric that is invariant under affine transformation. Such a metric is obtained as follows. The tangent space $T_{\Psi_0} PD(p)$ at each point $\Psi_0 \in PD(p)$ can be identified with $Sym(p)$, but we now equip $Sym(p)$ with the inner product $\langle Y_1,Y_2 \rangle_{\Psi_0}=\langle \Psi_0^{-1/2}Y_1\Psi_0^{-1/2},\Psi_0^{-1/2}Y_2\Psi_0^{-1/2} \rangle_{Sym(p)}=\mathrm{trace}(\Psi_0^{-1}Y_1\Psi_0^{-1}Y_2)$ which depends on the point $\Psi_0$ on the manifold where the tangent space is attached. For every pair $(\Psi_0,Y)\in PD(p) \times Sym(p)$, there is a unique geodesic passing through $\Psi_0$ with tangent vector $Y$ and it has expression $g(t)=\Psi_0^{1/2}\exp(t\Psi_0^{-1/2}Y\Psi_0^{-1/2})\Psi_0^{1/2}$, where $\exp(\cdot)$ denotes the matrix exponential and $t$ is the curve parameter. The corresponding geodesic distance between $\Psi_1,\Psi_2 \in PD(p)$ is then the affine-invariant metric  $d_R(\Psi_1,\Psi_2)=||\log(\Psi_1^{1/2}\Psi_2 \Psi_1^{1/2})||_{Sym(p)}$, with $\log(\cdot)$ being the logarithm matrix and $||\cdot||_{Sym(p)}$ the Frobenius norm. The exponential map $\exp_{\Psi_0}(Y)$ in $\Psi_0$ applied to $Y$ is the geodesic curve passing through $\Psi_0$ with tangent vector $Y$ evaluated for $t=1$, i.e.
$$
\exp_{\Psi_0}(Y)=\Psi_0^{1/2}\exp(\Psi_0^{-1/2}Y\Psi_0^{-1/2})\Psi_0^{1/2}
$$
and its inverse is the logarithmic map
$$
\log_{\Psi_0}(\Psi)=\Psi_0^{1/2}\log(\Psi_0^{-1/2}\Psi\Psi_0^{-1/2})\Psi_0^{1/2}.
$$

A complete description of the Riemannian geometry of $PD(p)$, with proofs of the above statements, can be found, e.g., in \citet{moakher2011riemannian}.

\medskip
We shall now illustrate the explicit expressions to apply the RDD-MK method discussed in Section \ref{sec:RDD-man}, when the Riemannian manifold $\mathcal{M}$ is $PD(p)$. The random observations $\boldchi_{\s_1},...,\boldchi_{\s_n}$ in $PD(p)$, at locations $\s_1,\ldots,\s_n$ in $D$, are assumed to be a partial observation of the random field $\{\boldchi_\s, \s \in D\}$ valued in $PD(p)$. In the $k$-th tile $D_k$, $k=1,...,K$, we represent the random field as
$$
\boldchi_\s(\Psi_k) = \exp_{\Psi_k}\{m_k+\bolddelta_\s\}= \Psi_k^{1/2}\exp(\Psi_k^{-1/2}(m_k+\bolddelta_\s)\Psi_k^{-1/2})\Psi_k^{1/2},
$$
with $\Psi_k \in PD(p)$ and $m_k,\bolddelta_\s \in Sym(p)$.

At each iteration of the bootstrap algorithm, the point $\Psi_k$ in the $k$-th tile is estimated as the sample Fr\'{e}chet mean of the observations within the $k$-th tile, i.e.,
\begin{equation}\label{eq:Psi_hat-PD}
\hat\Psi_k=\argmin{\Psi \in PD(p)} \sum_{\s_i \in D_k} \|\log(\Psi^{1/2}\boldchi_{\s_i}\Psi^{1/2})\|^2_{Sym(p)},
\end{equation}
Note that the sample Fr\'{e}chet mean is uniquely defined for a sample of observations belonging to $PD(p)$, as shown in \citet{moakher2005differential}. The stationary trace-variogram \eqref{eq:tr-var} for the transformed field $\log_{\Psi_k}(\boldchi_{\s_i})$ in this context reads
\begin{align*}
2\gamma(\s_i-\s_j;k)&=E[\| \log_{\Psi_k}(\boldchi_{\s_i}) - \log_{\Psi_k}(\boldchi_{\s_j})\|_{\Psi_k}^2]= \\
&=E[\mathrm{trace}((\Psi_k^{-1/2}(\log_{\Psi_k}(\boldchi_{\s_i}) - \log_{\Psi_k}(\boldchi_{\s_j}))\Psi_k^{-1/2})^2)].
\end{align*}
Its geographically-weighted estimator \eqref{eq:tr-var-est} is then found as
\begin{align}\label{eq:tr-var-est-PD}
\nonumber
\widehat{\gamma}(\h;k) &= \frac{\sum_{N(\h)} \mathcal{K}(c_k,\s_i)\mathcal{K}(c_k,\s_j) [\mathrm{trace}((\hat\Psi_k^{-1/2}(\log_{\hat\Psi_k}(\boldchi_{\s_i}) - \log_{\hat\Psi_k}(\boldchi_{\s_j}))\hat\Psi_k^{-1/2})^2)]}{2\sum_{N(\h)}
\mathcal{K}(c_k,\s_i)\mathcal{K}(c_k,\s_j)}=\\
& = \frac{\sum_{N(\h)} \mathcal{K}(c_k,\s_i)\mathcal{K}(c_k,\s_j) [\|\log_{\hat\Psi_k}(\boldchi_{\s_i}) - \log_{\hat\Psi_k}(\boldchi_{\s_j})\|_{\hat\Psi_k}^2]}{2\sum_{N(\h)}
	\mathcal{K}(c_k,\s_i)\mathcal{K}(c_k,\s_j)}.
\end{align}
where $\hat\Psi_k$ is defined in \eqref{eq:Psi_hat-PD}.
Finally, given the bootstrap kriging predictors $\{\boldchi^{*b}_{\s_0}\}_{b=1}^B$, the final RDD-MK predictor is found as their intrinsic mean in $PD(p)$, i.e.,
\begin{equation}\label{eq:aggregation-PD}
\boldchi^{*}_{\s_0}=\argmin{x \in PD(p)} \sum_{b=1}^B \|\log(x^{1/2}\boldchi^{*b}_{\s_0}x^{1/2})\|^2_{Sym(p)}.
\end{equation} 

\subsection{The space of correlation matrices (and the hypersphere)}\label{subsubsec:cor}

As a second example we consider the space of correlation matrices.
While in recent years there has been much interest in the statistical analysis of positive definite matrices \citep[see, e.g.,][]{Dryden2009, Yuan2012}, little attention has been devoted to the case of the space $\mathcal{R}$ of correlation matrices. These form a subset of the space of positive definite matrices, defined by the additional constraint of having ones on the diagonal. This additional constraint entails that it is not possible to use the geometry described in the previous subsection to deal with correlation matrices. For example, the geodesic in $PD(p)$ between two correlation matrices is not bounded to be in the space of correlation matrices and the local tangent approximation is not guaranteed to return valid correlation matrices.
Therefore, one needs to use a different space to carry out statistical analysis for correlation matrices.

It is possible to identify the space of full rank correlation matrices of dimension $p$ with a specific set of $p\times p$ upper triangular matrices, the \emph{Cholesky manifold}, whose properties have been studied by \citet{grubivsic2007efficient} in the context of rank reduction in the estimation of a correlation matrix.
The Cholesky manifold is defined as
$$
\text{Chol}(p)=\{H \in \mathbb{R}^{p\times p}: H_{ij}=0 \text{ if } j<i; \, H_{11}=1, \, ||H^{(i)}||_{\mathbb{R}^i}=1\},
$$
where $H^{(i)}$ denotes the vector of the first $i$ elements of the $i$-th column of $H$ (the remaining elements being $0$ by definition). This means that (i) if $H \in \text{Chol}(p)$,  the $i$-th column of $H$ can be identified with an element of the $i$-th dimensional hypersphere $S^i$ with radius $1$, and (ii) $\text{Chol}(p)$ is a product manifold with $\text{Chol}(p)=\mathbf{e}_1 \otimes \prod_{i=2}^p S^i$, $\mathbf{e}_1$ being the $p$-dimensional vector such that $e_{11}=1$ and $e_{1i}=0$ for $i=2,\ldots,p$. The geometry of the Cholesky manifold is therefore inherited from the geometry of the (hyper)spheres, since geodesic curves on the product manifold are tensor product of geodesic curves on the spheres.  In the following, we discuss how to apply the proposed method to the Cholesky manifold but the same tools can of course be used to analyse data belonging to a hypersphere.

The key motivation to introduce the Cholesky manifold is that it has  a one-to-one correspondence with the space of correlation matrices, since for any $H\in \text{Chol}(p)$, $R=H^TH$ is a correlation matrix and for any correlation matrix $R$, it exists $H\in \text{Chol}(p)$ such that $R=H^T H$ \citep[see][]{grubivsic2007efficient}. Moreover, since $R$ is a positive definite matrix, there is a unique Cholesky factorization such that $R=H^TH$, thus defining a bijection between the two spaces. In the following, we will denote with $Ch(R)$ the unique Cholesky factor such that $R=Ch(R)^T Ch(R)$. For example, in the case of $2\times 2$ correlation matrices,
$$
R=\left(\begin{array}{cc}
1 & \rho\\
\rho & 1\\
\end{array}
\right), \hspace{4mm}
H=Ch(R)=\left(\begin{array}{cc}
1 & \rho\\
0 & \sqrt{1-\rho^2}\\
\end{array}
\right). \hspace{4mm}
$$

We can therefore map the sample of correlation matrices to the Cholesky manifold, use the geometry of the manifold to carry out the analysis and then map back the predicted field to the space of correlation matrices.


\paragraph*{Riemannian structure of hyperspheres}
Let us first recall that the hypersphere $S^q=\{\mathbf{z} \in \mathbb{R}^q: ||\mathbf{z}||_{\mathbb{R}^q}=1\}$ can be equipped with a Riemannian manifold structure by considering it as a Riemannian submanifold of the embedding Euclidean space $\mathbb{R}^q$ with the usual inner product $\langle \mathbf{x},\mathbf{y} \rangle_{\mathbb{R}^q}=\mathbf{x}^T\mathbf{y}$, $\mathbf{x},\mathbf{y} \in \mathbb{R}^q$.
The tangent space to $S^q$ in $\mathbf{z}$ is $T_{\mathbf{z}} S^q=\{ \mathbf{y}\in \mathbb{R}^q: \langle \mathbf{z},\mathbf{y} \rangle_{\mathbb{R}^q}=0 \}$, i.e. the set of vector orthogonal to $\mathbf{z}$, which can be equipped with the usual inner product in $\mathbb{R}^q$. The geodesic curve passing through $\mathbf{z}_0 \in S^q$ in direction $\mathbf{y}\in T_{\mathbf{z}}S^q, \mathbf{y}\neq 0$ has expression
$$
\exp_{\mathbf{z}_0} (t \mathbf{y})=\cos(t ||\mathbf{y}||_{\mathbb{R}^q})\mathbf{z}_0+ \sin(t ||\mathbf{y}||_{\mathbb{R}^q})\frac{ \mathbf{y}}{||\mathbf{y}||_{\mathbb{R}^q}}.
$$
For $t=1$, we get the exponential map
$$
\exp_{\mathbf{z}_0} (\mathbf{y})=\cos(||\mathbf{y}||_{\mathbb{R}^q})\mathbf{z}_0+ \sin(||\mathbf{y}||_{\mathbb{R}^q})\frac{ \mathbf{y}}{||\mathbf{y}||_{\mathbb{R}^q}}.
$$
The geodesic distance between two elements $\mathbf{z}_0$ and  $\mathbf{z}$ in $S^q$ is the great circle distance
$$
d_{S^q}(\mathbf{z}_0,\mathbf{z})=\arccos(\langle\mathbf{z}_0, \mathbf{z} \rangle_{\mathbb{R}^q}).
$$
By inverting the exponential map, we obtain the logarithmic map
$$
\log_{\mathbf{z}_0}(\mathbf{z})= \frac{d_{S^q}(\mathbf{z}_0,\mathbf{z})}{||P_{\mathbf{z}_0}(\mathbf{z}-\mathbf{z}_0)||_{\mathbb{R}^q}} P_{\mathbf{z}_0} (\mathbf{z}-\mathbf{z}_0),
$$
where $P_{\mathbf{z}_0} (\mathbf{x})= \mathbf{x}- \langle \mathbf{z}_0,\mathbf{x} \rangle_{\mathbb{R}^q} \mathbf{z}_0$ is the projection of an element $\mathbf{x}\in \mathbb{R}^q$ into the tangent space $T_{\mathbf{z}_0} S^q$.

\paragraph*{Riemannian structure of $Chol(p)$}
We can extend the latter definitions to the Cholesky manifold by acting individually on each vector $H^{(i)}$ of the Cholesky matrix $H$, for $i=2,\ldots,p$. Let us define a map $\mathcal{R}: \mathbb{R} \otimes \bigotimes_{i=2}^p S^i \rightarrow \text{Chol}(p)$ such that
$$
\mathcal{R}(H^{(1)},H^{(2)},\ldots,H^{(p)})=H=\left\{
\begin{array}{ll}
H_{11}=H^{(1)}& \\
H_{ij}=0& \text{ if } i>j\\
H_{ij}=H^{(i)}_j & \text{ otherwise }.\\
\end{array}
\right.
$$
The tangent space in a point $H \in \text{Chol}(p)$ is $T_{H}\text{Chol}(p) = \bigotimes_{q=2}^p T_{H^{(q)}} S^q$, i.e., if $X \in T_{H}\text{Chol}(p)$, then $X$ is an upper triangular matrix $p \times p$ such that $X_{11}=0$ and $X^{(q)}\in T_{H^{(q)}} S^q$ for $q=2,\ldots,p$, where $X^{(q)}$ again denotes the vector of the first $q$ elements of the $q$-th column of $X$. The geodesics on $\text{Chol}(p)$ passing through $H$ in direction $X$ have the form
$$
\exp_{H}(tX)=\mathcal{R}\big(1,\exp_{H^{(2)}}(t X^{(2)}),\ldots, \exp_{H^{(p)}}(t X^{(p)})\big),
$$
where $\exp_{H^{(q)}}(t \mathbf{x}_q)$ denotes the geodesic on $S^q$.
The exponential and logarithmic maps are therefore
$$
\exp_{H}(X)=\mathcal{R}\big(1,\exp_{H^{(2)}}(X^{(2)}),\ldots, \exp_{H^{(p)}}(X^{(p)})\big),
$$
and
$$
\log_{H}(Z)= \mathcal{R}\big(0,\log_{H^{(2)}}(Z^{(2)}),\ldots, \log_{H^{(p)}}(Z^{(p)})\big)
$$
respectively, $\log_{H^{(q)}}(Z^{(q)})$ denoting the logarithmic map in $S^q$ and $H^{(q)},Z^{(q)}$ being the vectors of the first $q$ elements of the $q$-th column of $H$ and $Z$. The geodesic distance between $H$ and $Z$ in $\text{Chol}(p)$ is
\begin{equation}\label{eq:dist-cor}
d_{\text{Chol}(p)}(H,Z)=\sqrt{\sum_{q=2}^p d_{S^q}^2(H^{(q)},Z^{(q)})}.
\end{equation}

\paragraph*{RDD-MK with correlation matrices}
The RDD-MK algorithm in the case of correlation matrices works as follow. The random correlation matrix $\boldzeta_\s$ at location $\s\in D_k$ is decomposed in the random Cholesky factors $\boldzeta_\s = \boldchi_\s^T\boldchi_\s$, $\boldchi_\s \in Chol(p) $. Each factor is modelled in the $k$-th tile as
$$
\boldchi_\s(\Psi_k) = \exp_{\Psi_k}\{m_k+\bolddelta_\s\},$$
where $\exp_{\Psi_k}(\cdot)$ denotes the exponential map on the Cholesky manifold in $\Psi_k$ and $m_k,\bolddelta_\s \in T_{\Psi_k}\text{Chol}(p)$.

At each iteration of the bootstrap algorithm, the point $\Psi_k$ in the $k$-th tile can be estimated as the intrinsic mean of the observations within the $k$-th tile, i.e.
$$
\hat\Psi_k=\argmin{\Psi \in Chol(p)} \sum_{\s_i \in D_k} d^2_{\text{Chol}(p)}(\Psi,\boldchi_{\s_i}),
$$
where $d^2_{\text{Chol}(p)}(\cdot,\cdot)$ is defined in \eqref{eq:dist-cor}.

However, for the case of (hyper)spheres -- and, as a consequence, of the Cholesky manifold -- unicity of the intrinsic mean is guaranteed only if the support of the distribution is not too large. While this can in principle be obtained by considering a smaller tile, the price to pay on the accuracy in the estimation of the spatial dependence may be too high. In this case, it is possible to use the sample extrinsic mean as tangent point $\Psi_k$. This is obtained by computing the arithmetic mean for the column vectors and then projecting them on the sphere, i.e., rescaling each column vector to have norm $1$.

The stationary trace-variogram \eqref{eq:tr-var} for the transformed field $\log_{\Psi_k}(\boldchi_{\s}) \in T_{\Psi_k}\text{Chol}(p)$  is then
\begin{align*}	
2\gamma(\s_i-\s_j;k)=&E[\| \log_{\Psi_k}(\boldchi_{\s_i}) -\log_{\Psi_k}(\boldchi_{\s_j})\|^2_{T_{\Psi_k}\text{Chol}(p)}]=\\
=&E\left[\sum_{q=2}^{p} \| (\log_{\Psi_k}(\boldchi_{\s_i}) -\log_{\Psi_k}(\boldchi_{\s_j}))^{(q)}\|^2_{\mathbb{R}^q}\right],
\end{align*}
where $ \|x\|^2_{T_{\Psi_k}\text{Chol}(p)}=\langle x, x\rangle_{T_{\Psi_k}\text{Chol}(p)}$. Similarly, its empirical estimator in \eqref{eq:tr-var-est} reads
\begin{equation*}\label{eq:tr-var-est-chol}
\widehat{\gamma}(\h;k) = \frac{\sum_{N(\h)} \mathcal{K}(c_k,\s_i)\mathcal{K}(c_k,\s_j) \| \log_{\Psi_k}(\boldchi_{\s_i}) -\log_{\Psi_k}(\boldchi_{\s_j})\|^2_{T_{\Psi_k}\text{Chol}(p)}}{2\sum_{N(\h)}
\mathcal{K}(c_k,\s_i)\mathcal{K}(c_k,\s_j)}, \quad\h\in\mathbb{R}^d.
\end{equation*}

Given the bootstrap kriging predictors $\{\boldchi^{*b}_{\s_0}\}_{b=1}^B$, the final RDD-MK predictor is found as their intrinsic mean as \begin{equation*}\label{eq:aggregation-PD}
\boldchi^{*}_{\s_0}=\argmin{x \in Chol(p)} \sum_{b=1}^B d^2_{\text{Chol}(p)}(x,\boldchi^{*b}_{\s_0})
\end{equation*}
and the corresponding correlation matrix is $R_{\s_0}^*=(\boldchi^{*}_{\s_0})^T \boldchi^{*}_{\s_0}$.

\section{Two simulation studies}\label{sec:simu}
\subsection{Covariance matrices over a C-shaped domain}\label{subsec:covsim}
We here illustrate a simulation study to test the performances of the RDD-MK methodology described in Section \ref{sec:RDD-man}, in the case of $PD(p)$, with $p=2$.

\begin{figure}
  \centering
  \includegraphics[width=.7\textwidth]{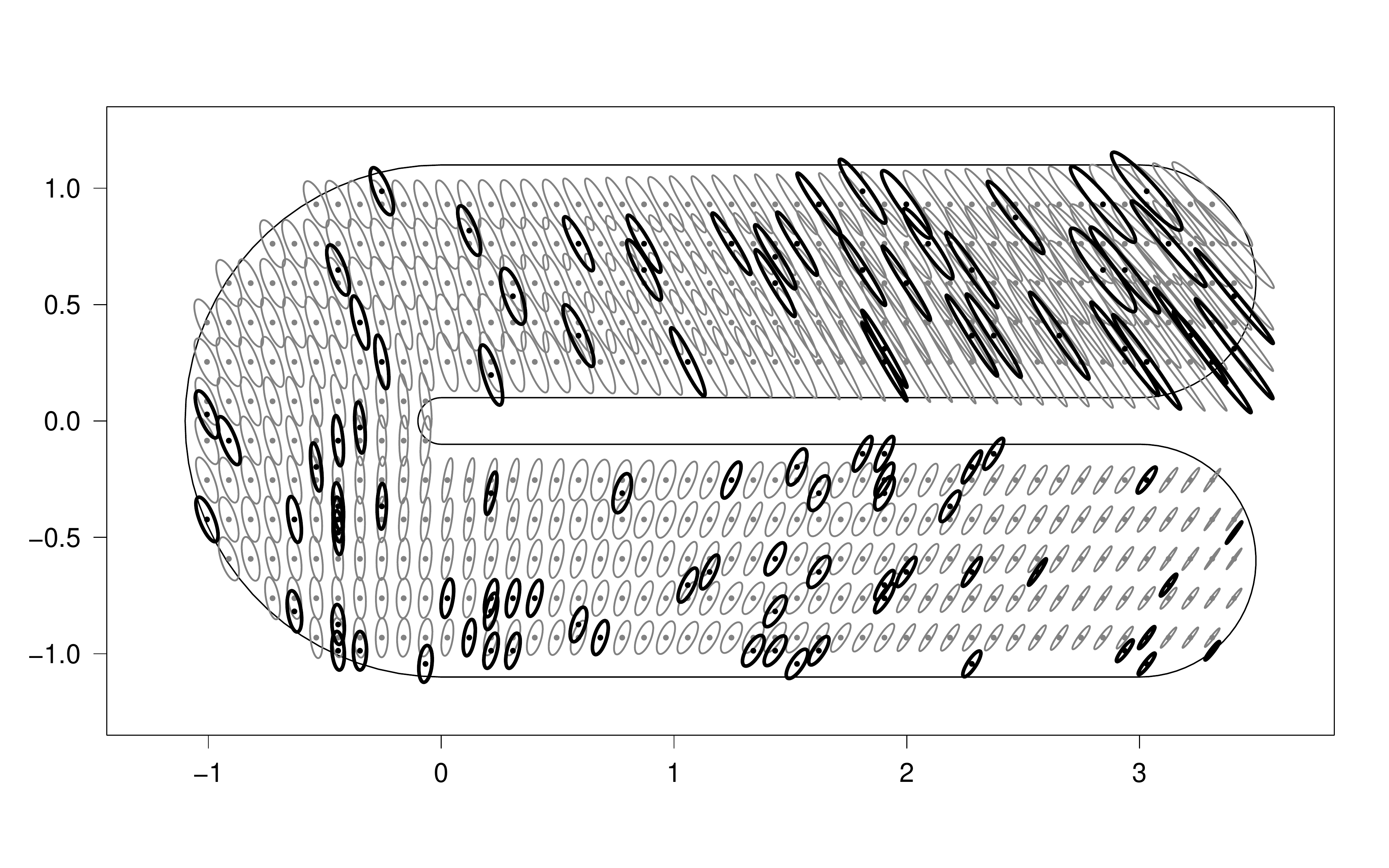}\\
  \caption{Realization of a locally stationary random field in $PD(2)$ on a C-shaped domain. Positive definite matrices are represented through the associated quadratic form. Grey ellipses represent a sub-sample of the reference realization, thick black ellipses represent the sampled observations. }\label{fig:Cdomain-data}
\end{figure}

The spatial domain we consider is the C-shaped domain of \citet[][and references therein]{Ramsay2002,WoodEtAl2008,SangalliEtAl2013, MenafoglioEtAl2018}, which is represented in Figure \ref{fig:Cdomain-data}. Inspired by the non-stationary field considered by these authors, we generate a random realization of a field in $PD(2)$ over the C by following three steps, namely (i) construction of a grid of tangent points $\Psi_\s$, (ii) construction of a zero-mean locally stationary process $\{\bolddelta_\s, \s\in D\}$ and a drift $A(\s)$ in $Sym(p)$, and (iii) combination of the results to build $\{\boldchi_\s, \s\in D\}$ as
\begin{equation}\label{eq:sim-model}
\boldchi_s = \exp_{\Psi_\s}(A(\s)+\bolddelta_\s), \quad \s \in D.
\end{equation}

In all these steps, the construction is made on a fine grid over a rectangular planar domain $(\varphi, r)$, and then the planar coordinates are mapped into the C by using $(\varphi,r)$ as centerline and radius coordinates along the C, finally obtaining a grid $\mathcal{G}$ of 1582 points on the C domain. Although this generating procedure might be seen as an \emph{ad hoc} technique for this domain, generating random fields on non-Euclidean domains is still an open challenge in geostatistics, due to the absence of valid variogram models (see, e.g., \citet{Curriero2006}). Note that the dimensions of the rectangle (i.e., the ranges for $\varphi,r$) are set as to obtain a C inscribed in the cartesian rectangle $[-1,3.5]\times[-1.1,1.1]$.

The elements $\Psi_\s$ are here built as $\Psi_\s = 0.5\cdot\alpha(\s)\cdot\exp_{\Sigma}\{A(\s)\}$, with
\begin{align}
\nonumber&\Sigma = \left(
           \begin{array}{cc}
             2 & 1 \\
             1 & 2 \\
           \end{array}
         \right)\\
\label{eq:A-drift}&A(\s)  = \left(
           \begin{array}{cc}
             0.5 & 0.4 \\
             0.4 & 0.5 \\
           \end{array}
         \right)\phi + \left(
           \begin{array}{cc}
             0.2 & -0.1 \\
             -0.1 & 0.2 \\
           \end{array}
         \right)\cdot(\phi_M - \phi)
         + \left(
           \begin{array}{cc}
             -0.2 & 0.1 \\
             0.1 & 0.4 \\
           \end{array}
         \right)\cdot r,\\
\nonumber& \alpha(\s) = \left(0.1 + \frac{\varphi_M-\varphi}{\varphi_M}\right)^{1/2}
\end{align}
for $\s = (\varphi, r)$, $\varphi \in [0, \varphi_M]$, $\varphi_M = 8.88$, and $r\in [-0.5, 0.5]$.
The locally stationary random field in $Sym(2)$, $\{\bolddelta_\s, \s\in D\}$, is generated by simulating its matrix elements $\bolddelta_\s^{(ij)}$, $i,j = 1,2$ as follows. First, the elements $\widetilde{\bolddelta}_\s^{(11)}$, $\widetilde{\bolddelta}_\s^{(12)}$, $\widetilde{\bolddelta}_\s^{(22)}$ are sampled from three independent zero-mean Gaussian stationary random fields, with spherical variogram of range $\rho=10$ and sill $\sigma^2 = (3.75)^2$, and by symmetry $\widetilde{\bolddelta}_\s^{(21)} = \widetilde{\bolddelta}_\s^{(12)}$. The elements $\bolddelta_\s^{(ij)}$, $i,j=1,2$, are then built as $\bolddelta_\s^{(ij)}=\alpha(s)^2\widetilde{\bolddelta}_\s^{(ij)}$. Note that the latter construction entails a non-stationary covariance model. Indeed, the process variance slowly varies with $\s$, according to $\alpha(s)^2$.
Finally the random process is built as in \eqref{eq:sim-model}, by using the drift $A(\s)$ in \eqref{eq:A-drift}. The results are represented as grey ellipses in Figure \ref{fig:Cdomain-data}, where matrices are represented through the ellipses generated by the associated quadratic forms. Here, the direction of the axes were determined as the matrix eigenvectors, the length of the semi-axes set proportional to the square-root of the eigenvalues, and the ellipses' radius fixed to 0.02 for representation purposes.

To test the performances of the RDD-MK method, we randomly draw $n=100$ locations, uniformly in the C-shaped domain. The selected observations are depicted as thick black ellipses in Figure \ref{fig:Cdomain-data}. Graphical inspection of Figure \ref{fig:Cdomain-data} suggests the existence of an evident trend in the data, which follows the shape of the C. Indeed, the model parameters slowly vary as a function of the centerline and of the radius of the C. However, an abrupt discontinuity would be observed if moving from one branch to the other through the boundaries of the C. This motivates the use of a non-Euclidean distance to build the Voronoi tessellations \eqref{eq:voronoi}, similarly as in \citep{MenafoglioEtAl2018}.

\begin{figure}
  \centering
  \includegraphics[width=.7\textwidth]{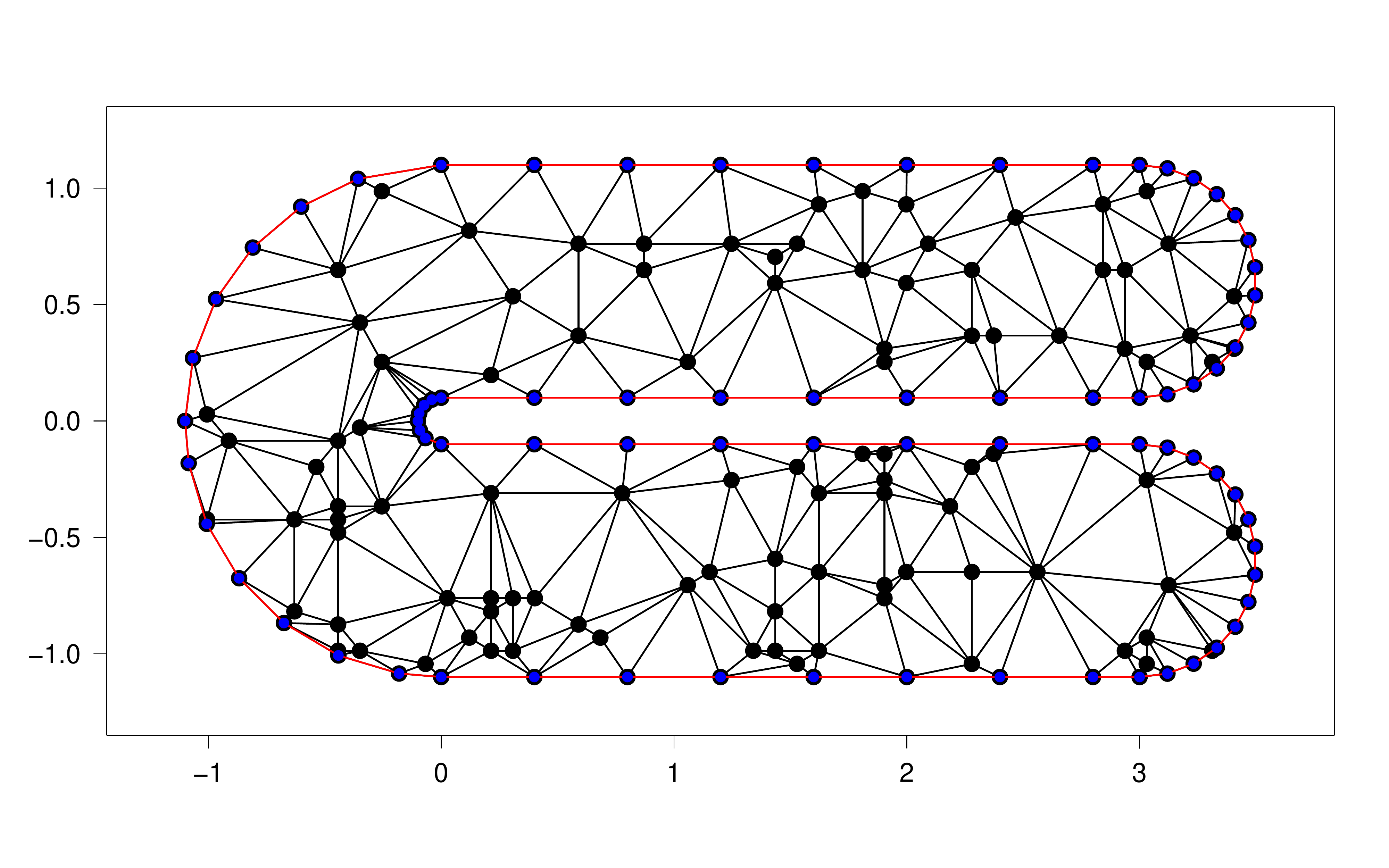}\\
  \caption{Representation of the C-shaped domain through a Delaunay triangulation. Black points indicate data locations, blue points indicate the point used to define the domain border.}\label{fig:Cdomain}
\end{figure}

To compute the latter distance, the non-Euclidean metric in the C is here approximated through the graph-distance (shortest path on the graph) among locations induced by a Delaunay triangulation \citep{HjelleDaehlen2006} with vertices in the data locations (Figure \ref{fig:Cdomain}). The finer the Delaunay triangulation, the better the graph distance approximates the geodesic distance between points within the C. A similar idea is used in \citep{MenafoglioEtAl2018}.

For the variogram estimation, we consider a Gaussian kernel $\mathcal{K}_\epsilon$, with bandwidth $\epsilon = 1.5$. The latter parameter is set as to balance the trade-off between the locality of the results and the stability of the algorithm. A spherical model with nugget is considered as a parametric valid model for the variogram. The number of bootstrap iteration is set to $B=100$ and the prediction performances tested for values of $K$ in $\{1,2,4,6,8,10\}$. Note that the case $K=1$ coincides with the model of \citep{PigoliEtAl2016}. 

\begin{figure}
  \centering
  \subfigure[$K=1$]{\includegraphics[width=.475\textwidth]{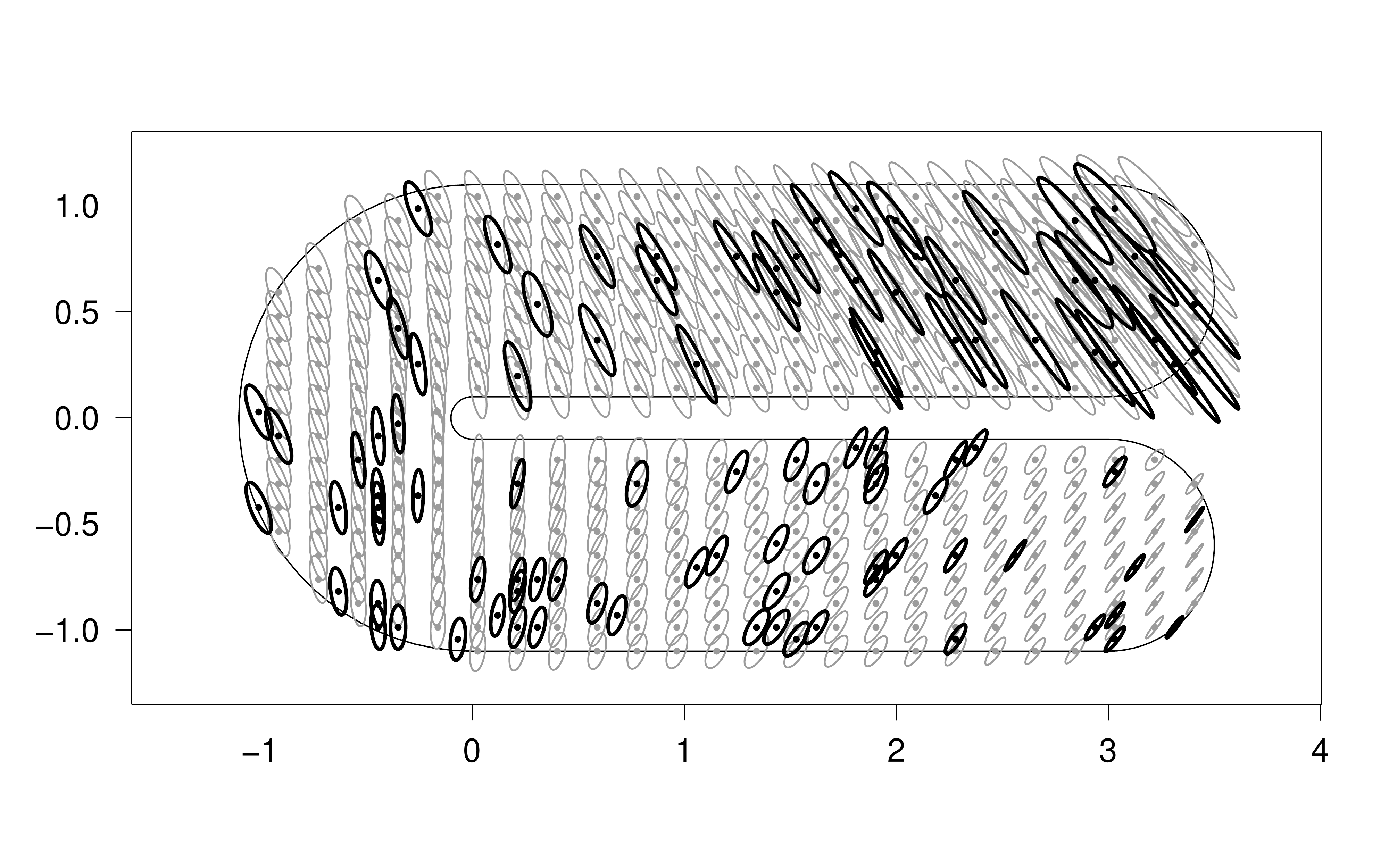}}
  \subfigure[$K=2$]{\includegraphics[width=.475\textwidth]{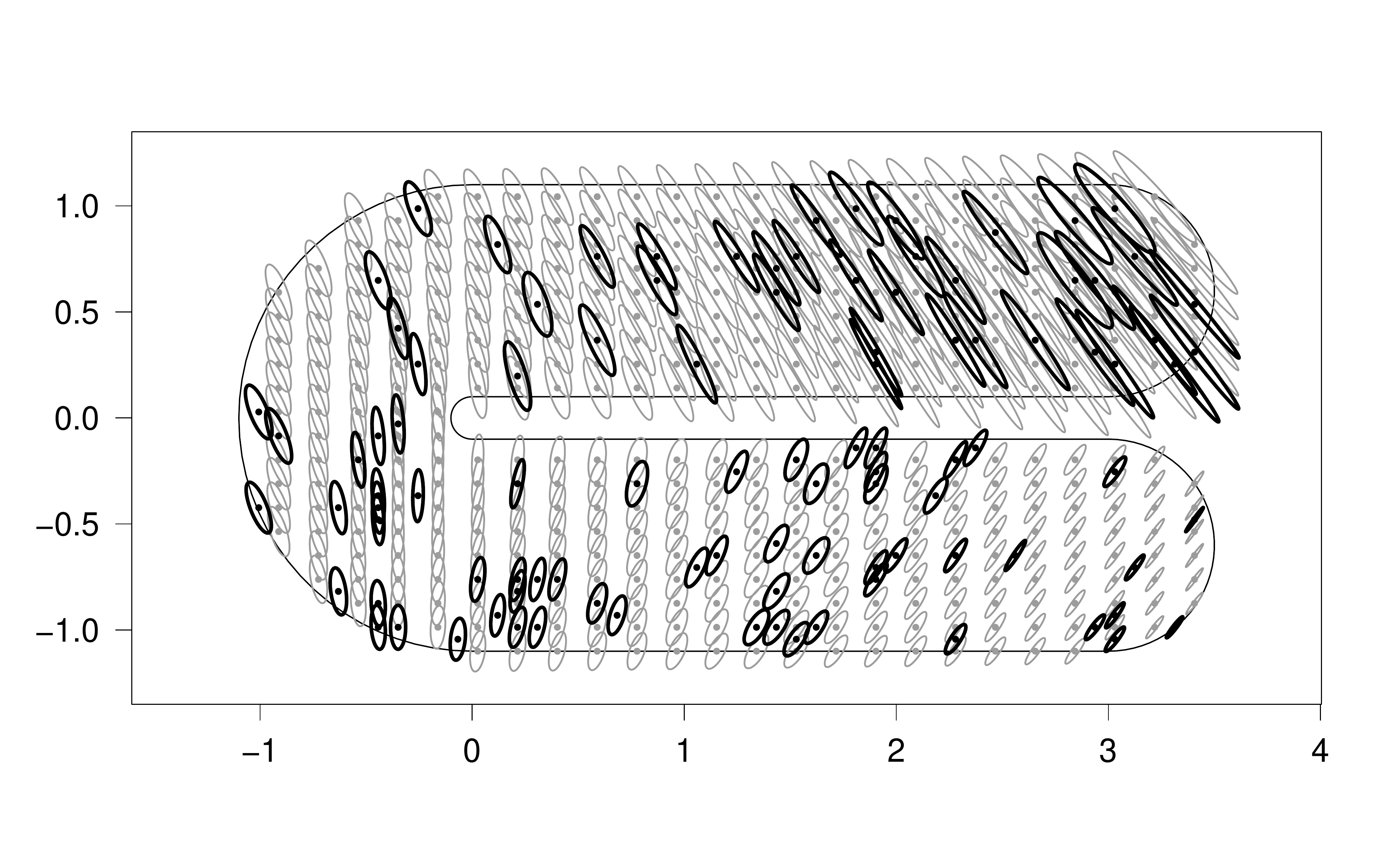}}
  \subfigure[$K=4$]{\includegraphics[width=.475\textwidth]{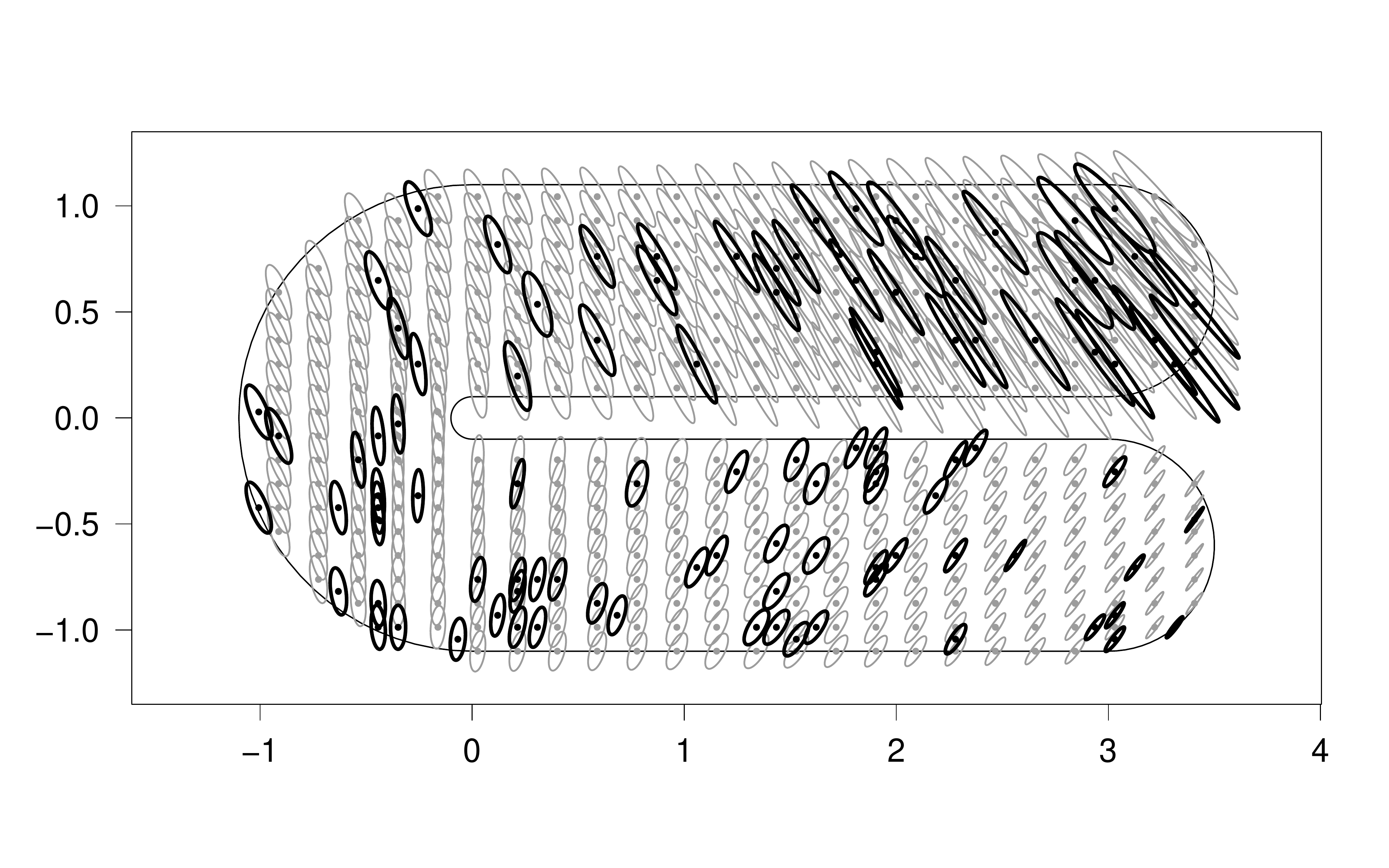}}
  \caption{Spatial prediction via RDD-MK. Predicted covariance matrices are represented as ellipses. Predictions refer to the same grid as the reference realization in Figure \ref{fig:Cdomain-data}.}\label{fig:results-cov}
\end{figure}

\begin{figure}
  \centering
  \subfigure[$K=1$]{\includegraphics[width=.475\textwidth]{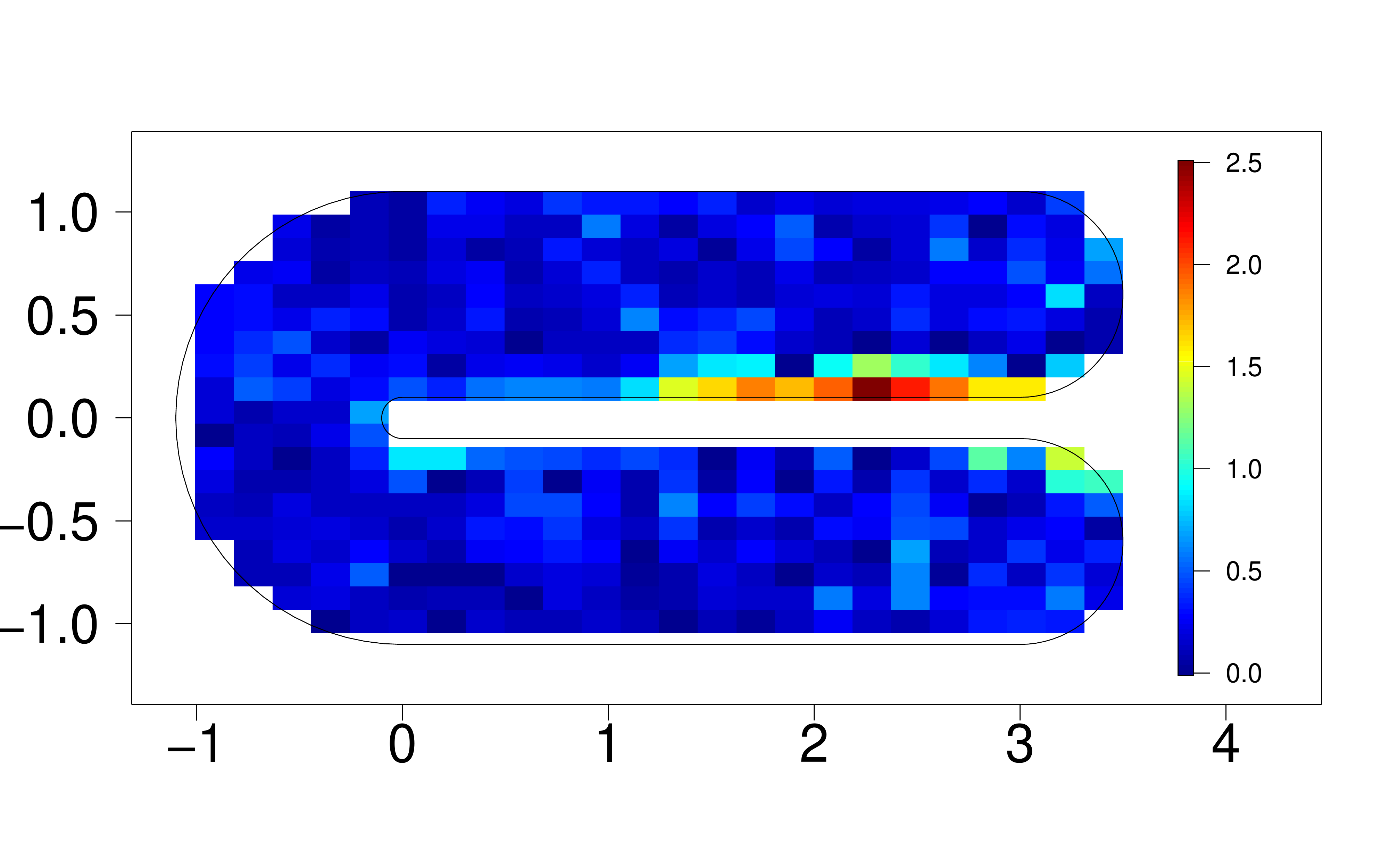}}
  \subfigure[$K=2$]{\includegraphics[width=.475\textwidth]{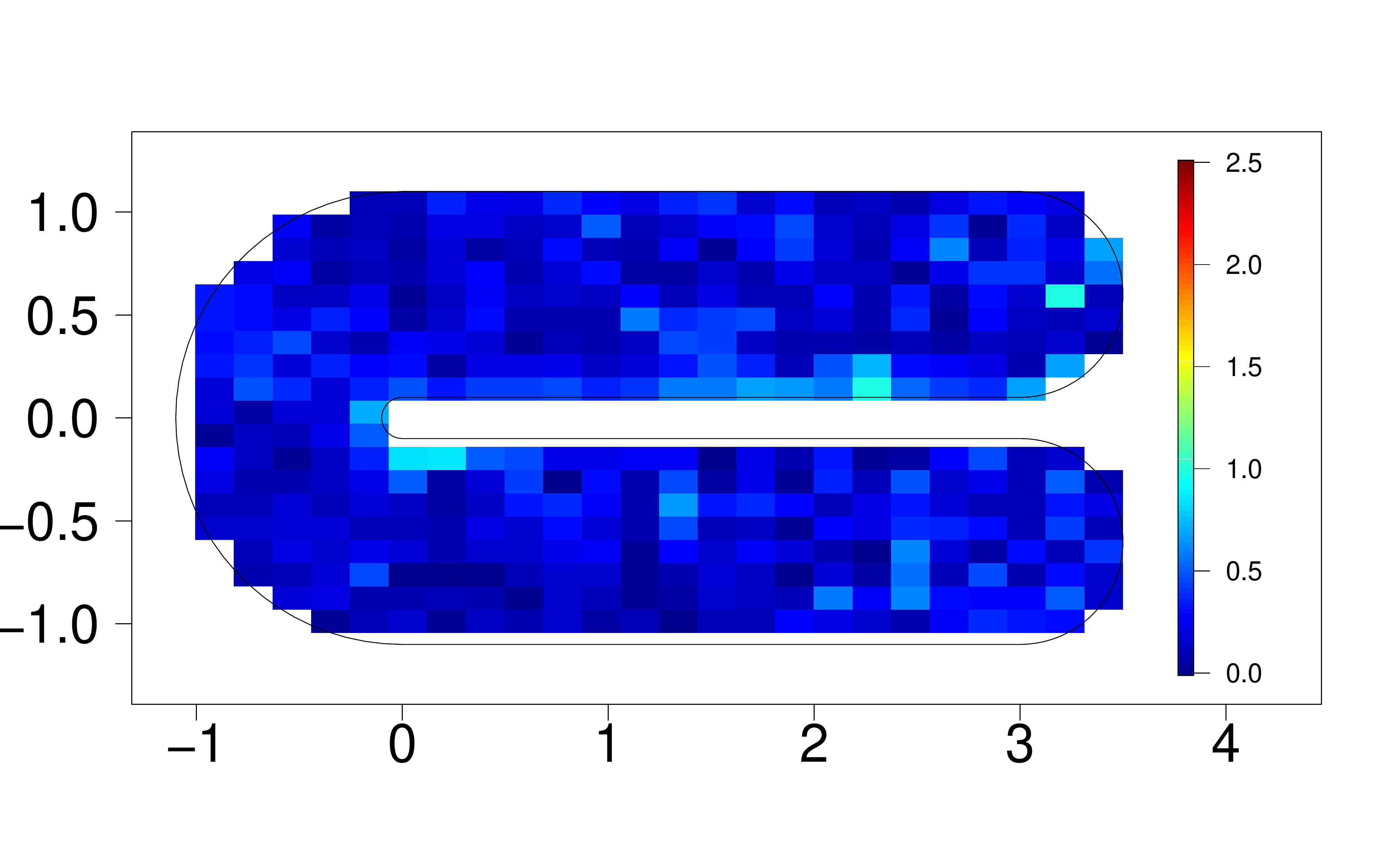}}
  \subfigure[$K=4$]{\includegraphics[width=.475\textwidth]{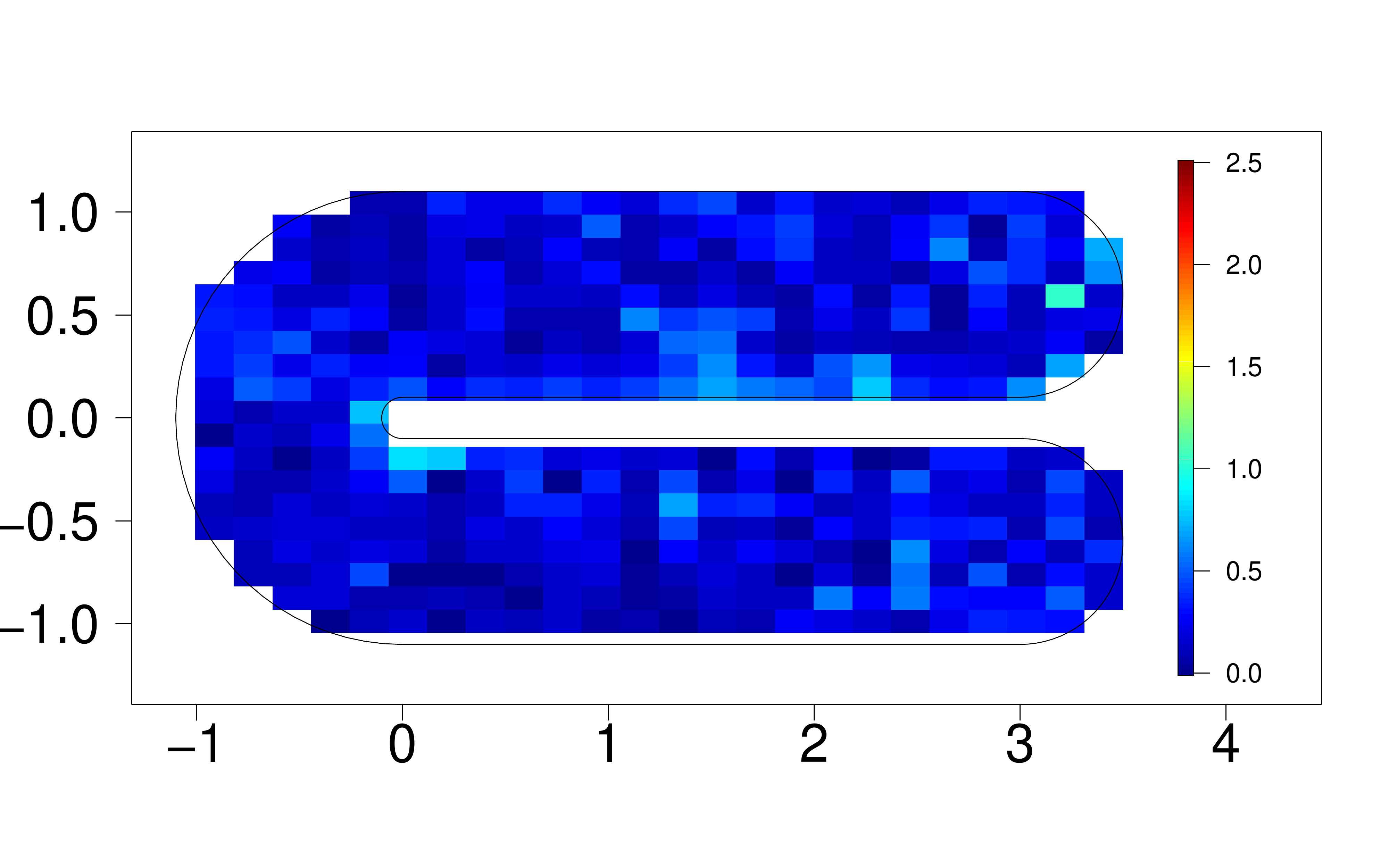}}
  \caption{Prediction error of RDD-MK method, computed as Riemannian distance between prediction and target, $d_R(\boldchi_{\s_0}, \boldchi^*_{\s_0})$.}\label{fig:cov-errors}
\end{figure}

The results of the RDD-MK algorithm for $K=1,2,4$ are shown in Figure \ref{fig:results-cov}, which displays the predictions obtained on the same grid $\mathcal{G}$ used for generating the data. Figure \ref{fig:cov-errors} reports the errors with respect to the reference realization, computed as $d_R(\boldchi_{\s_0}, \boldchi^*_{\s_0})$, $\boldchi_{\s_0}$ denoting the reference realization at $\s_0\in \mathcal{G}$. Both figures suggest that a significant improvement is obtained when moving from $K=1$ to $K=2$, supporting the use of an RDD algorithm. Indeed, when $K=1$ the method cannot account for the non-stationary behavior of the data, particularly at the boundary separating the two branches of the C. Here, increasing $K$ enables to better account for the shape of the domain, avoiding to incorrectly combine the information in the different parts of the domain.


\begin{figure}
  \centering
  \includegraphics[width=.49\textwidth]{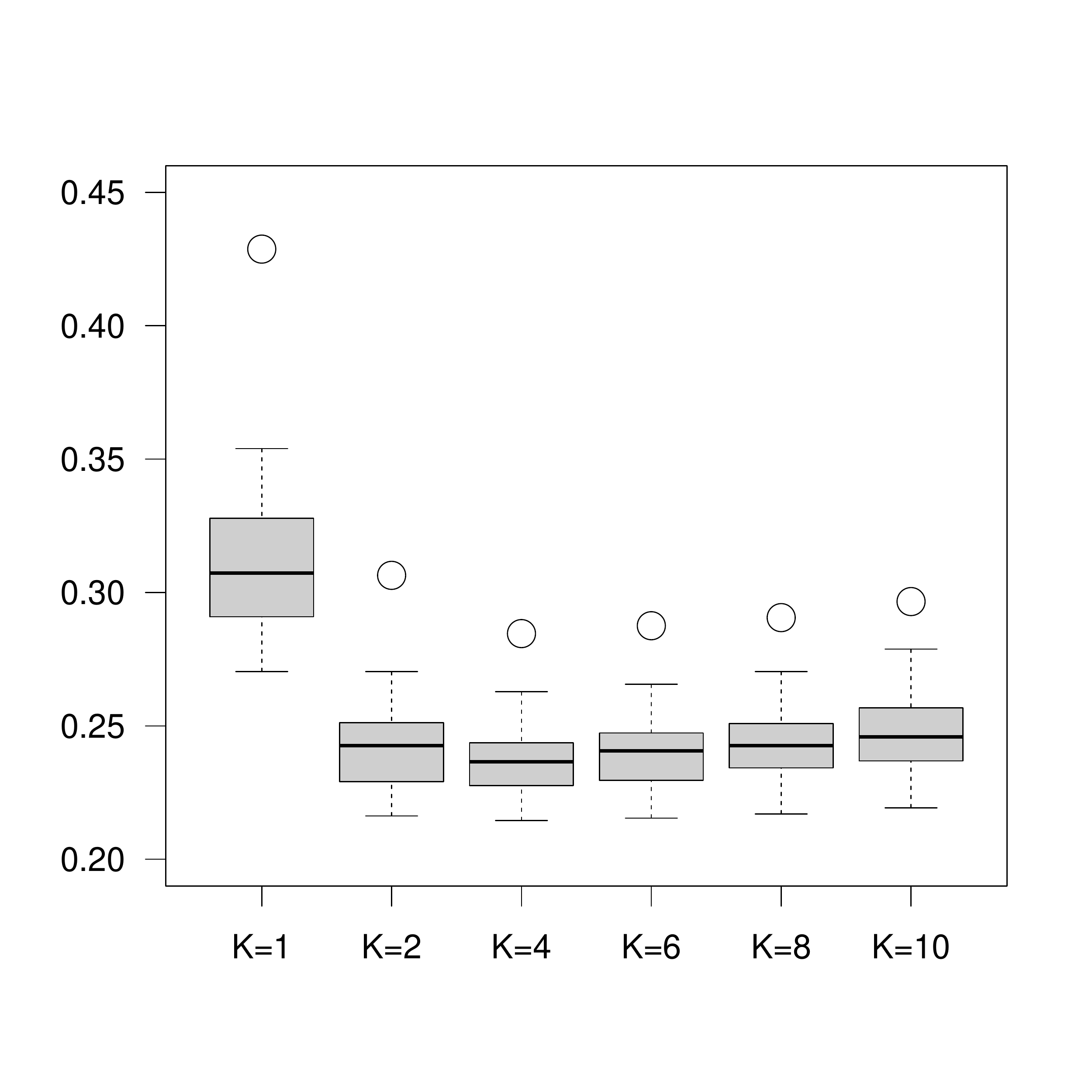}
  \caption{Monte Carlo study on covariance matrices. Boxplots of the average mean square prediction error, $K\in\{1,2,4,6,8,10\}$.}\label{fig:boxplot-cov}
\end{figure}

\medskip
To further test the prediction performances of RDD-MK, we perform a Monte Carlo study. For 30 subsamples randomly drawn from the realization depicted in Figure \ref{fig:Cdomain-data}, we repeat the same procedure detailed before, i.e., domain representation and RDD-MK prediction with the same parameter settings. For the $j$-th subsample, $j=1,2,...,30$, the prediction error is computed on the grid, i.e., $SPE(\s_0,j,K) = d_R(\boldchi_{\s_0}, \boldchi^{*(j,K)}_{\s_0})$, $\s_0\in \mathcal{G}$, $K\in\{1,2,4,8,10\}$. Each error map is then summarized by its average $MSPE(j,K) = \sum_{\s_0\in\mathcal{G}} d_R(\boldchi_{\s_0}, \boldchi^{*(j,K)}_{\s_0})/|\mathcal{G}|$, $|\mathcal{G}|$ denoting the cardinality of $\mathcal{G}$.  Figure \ref{fig:boxplot-cov} reports the boxplot of $MSPE(\cdot, K)$, for $K\in\{1,2,4,6,8,10\}$, whereas Table \ref{tab:sim-cov} reports its mean, median and standard deviation. These results confirm a clear improvement of the prediction performances when using RDD-MK ($K\geq 2$) with respect to the stationary case of \citep{PigoliEtAl2016} ($K=1$). Although a clear elbow is visible at $K=2$, the best results -- in terms of MSPE -- are obtained for $K=4$. No improvement is attained by a further increase of $K$, due to the bias-variance trade-off affecting the choice of this parameter.

\begin{table}
  \centering

\begin{tabular}{ccccccc}
  \toprule
  $K$   & 1 & 2 & 4 & 6 & 8 & 10 \\\midrule
  Mean  & 0.3127 & 0.2433 & \textbf{0.2387} & 0.2413 & 0.2444 & 0.2482\\
  Median& 0.3073 & 0.2426 & \textbf{0.2365} & 0.2407 & 0.2427 & 0.2459\\
  SD    & 0.0316 & 0.0180 & \textbf{0.0156} & 0.0156 & 0.0157 & 0.0167\\
  \bottomrule
\end{tabular}
   \caption{Monte Carlo study on covariance matrices. Mean, median and standard deviation of the average mean square prediction error, for $K\in\{1,2,4,6,8,10\}$. The minimum over $K$ is highlighted in bold.}\label{tab:sim-cov}
\end{table} 

\subsection{Correlation matrices over a C-shaped domain}\label{subsec:corsim}
\begin{figure}
	\centering
  \subfigure[$K=1$]{\includegraphics[width=.475\textwidth]{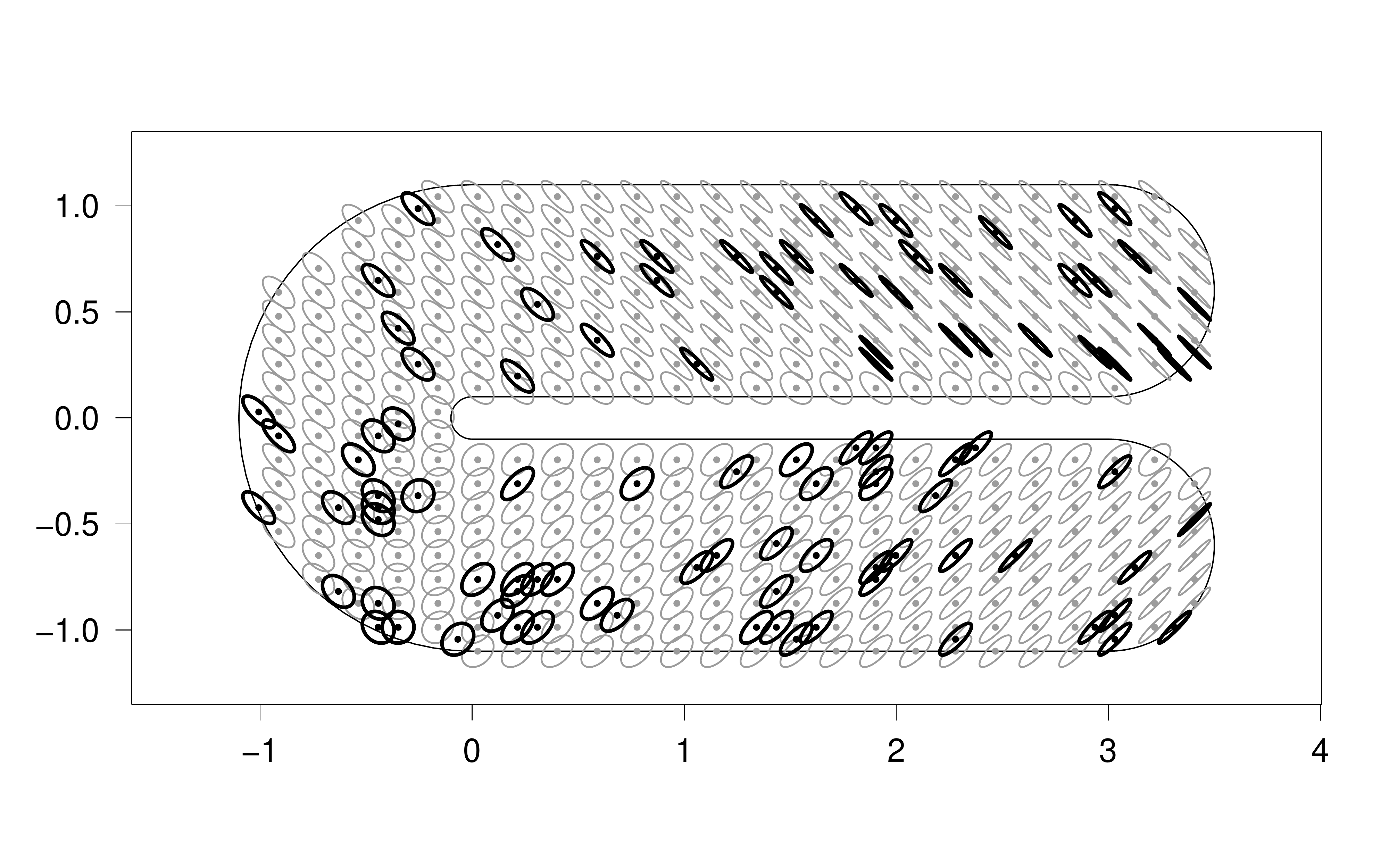}}
  \subfigure[$K=2$]{\includegraphics[width=.475\textwidth]{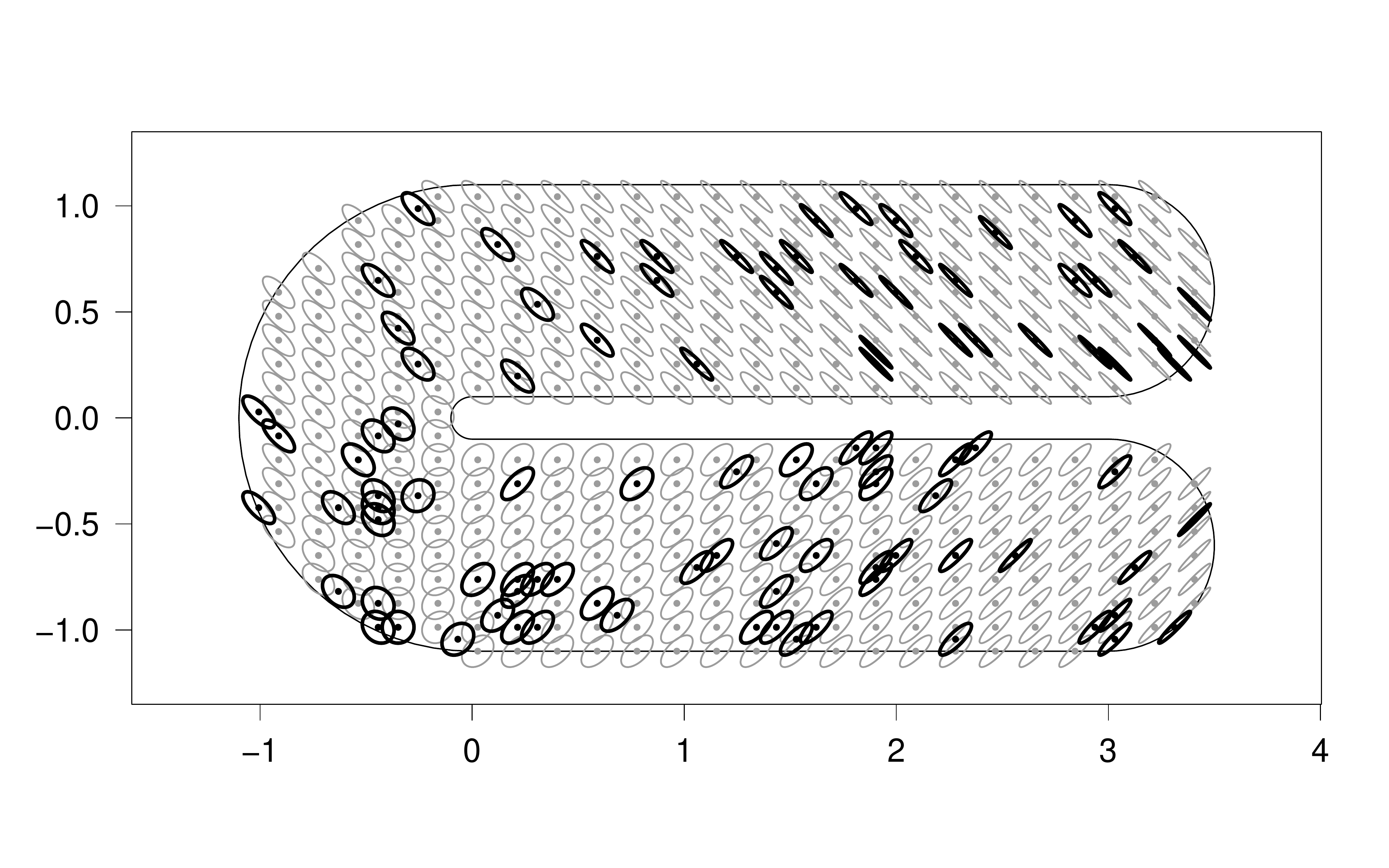}}
  \subfigure[$K=4$]{\includegraphics[width=.475\textwidth]{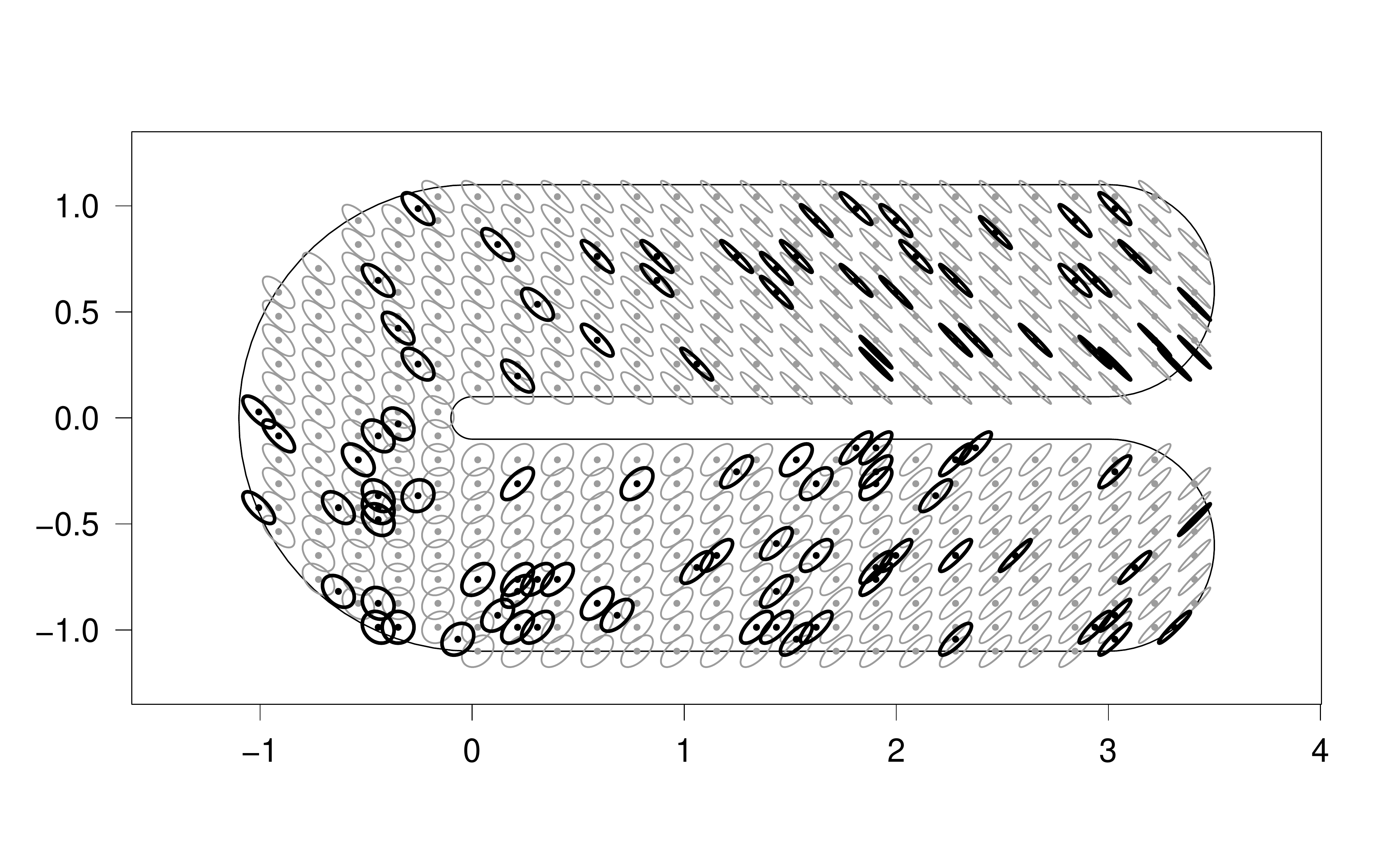}}
	\caption{Spatial prediction of correlation matrices via RDD-MK. Predicted correlation matrices are represented as ellipses. Predictions refer to the same grid as the reference realization in Figure \ref{fig:Cdomain-data}.}\label{fig:results-cor}
\end{figure}

\begin{figure}
	\centering
  \subfigure[$K=1$]{\includegraphics[width=.475\textwidth]{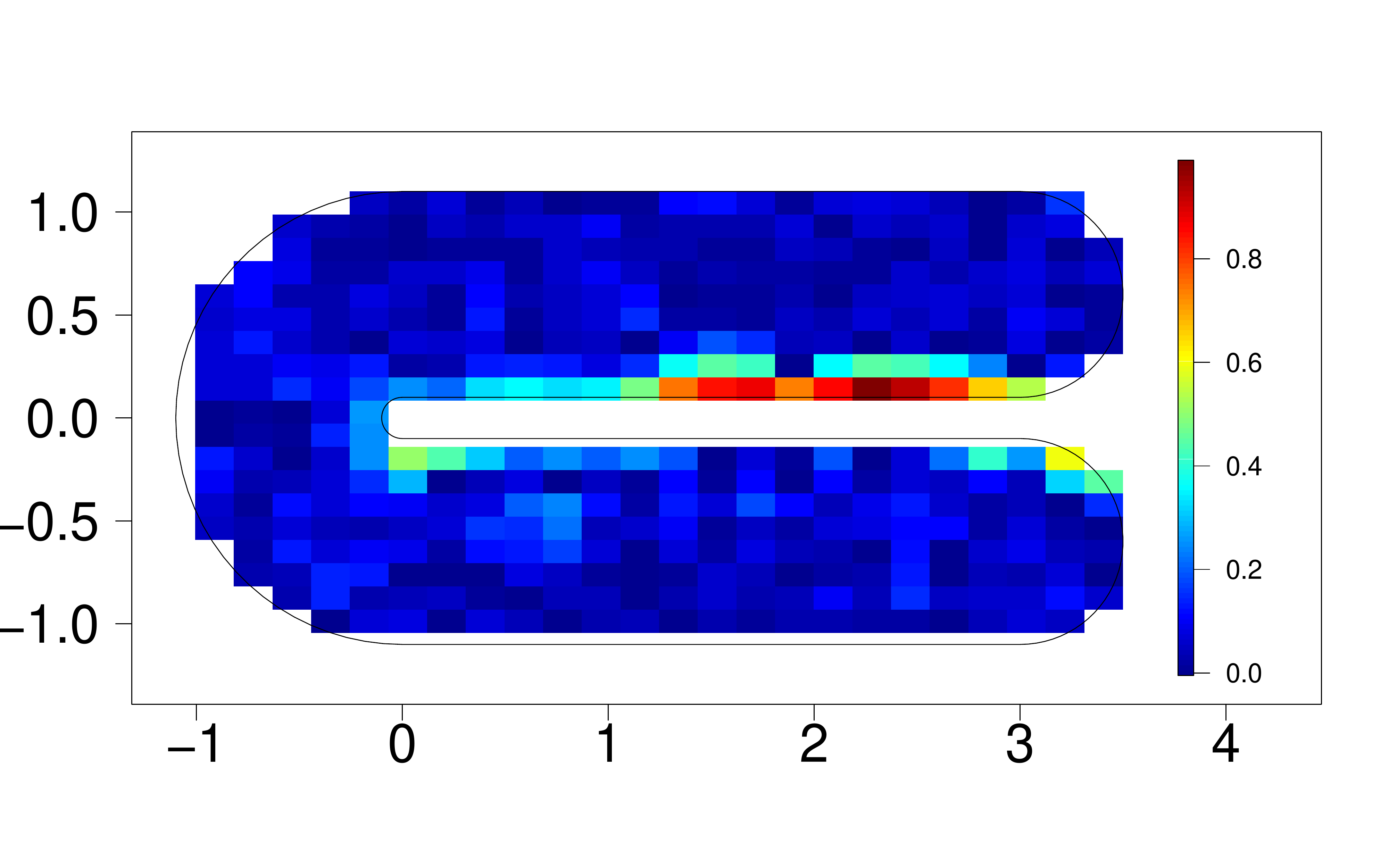}}
  \subfigure[$K=2$]{\includegraphics[width=.475\textwidth]{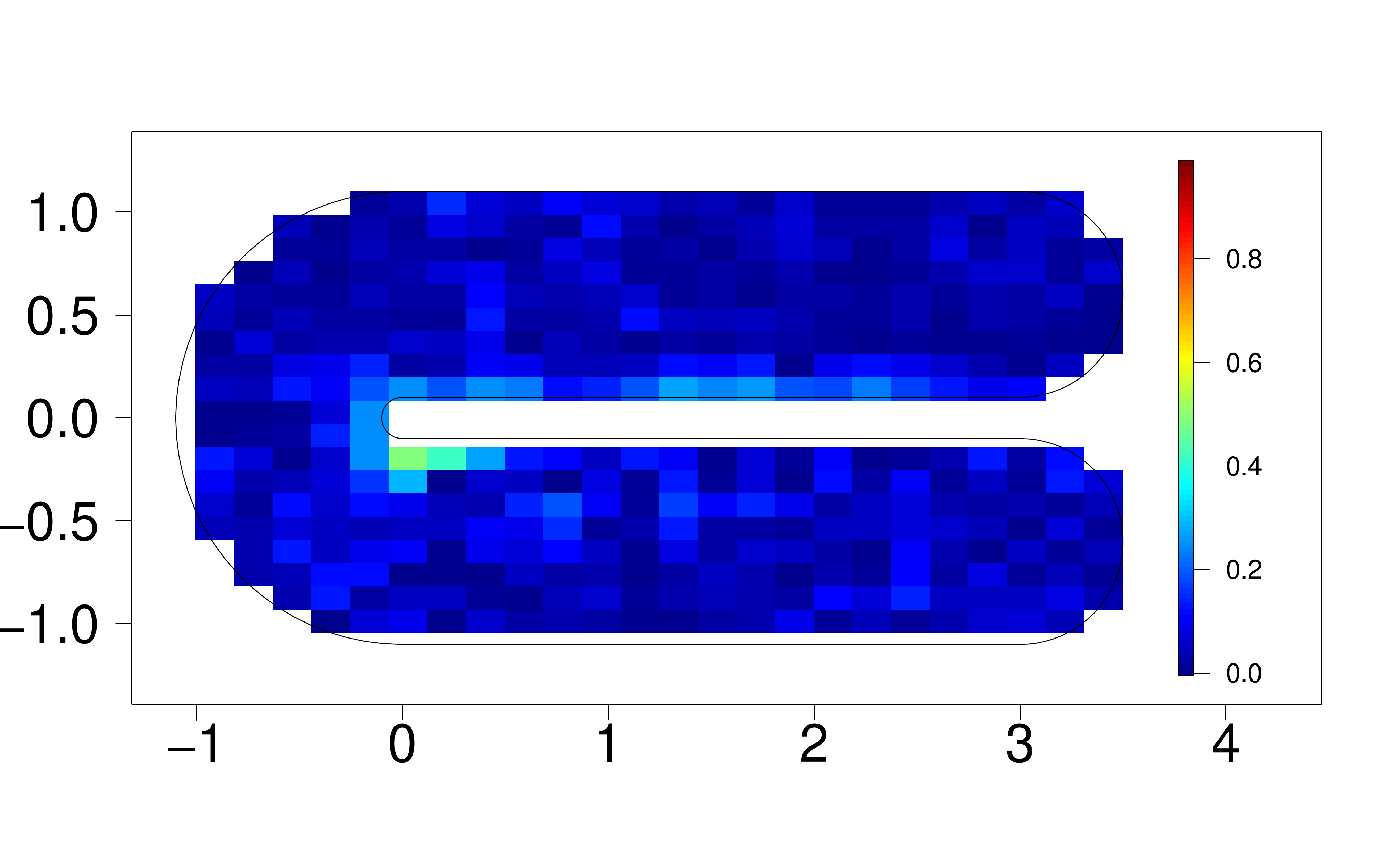}}
  \subfigure[$K=4$]{\includegraphics[width=.475\textwidth]{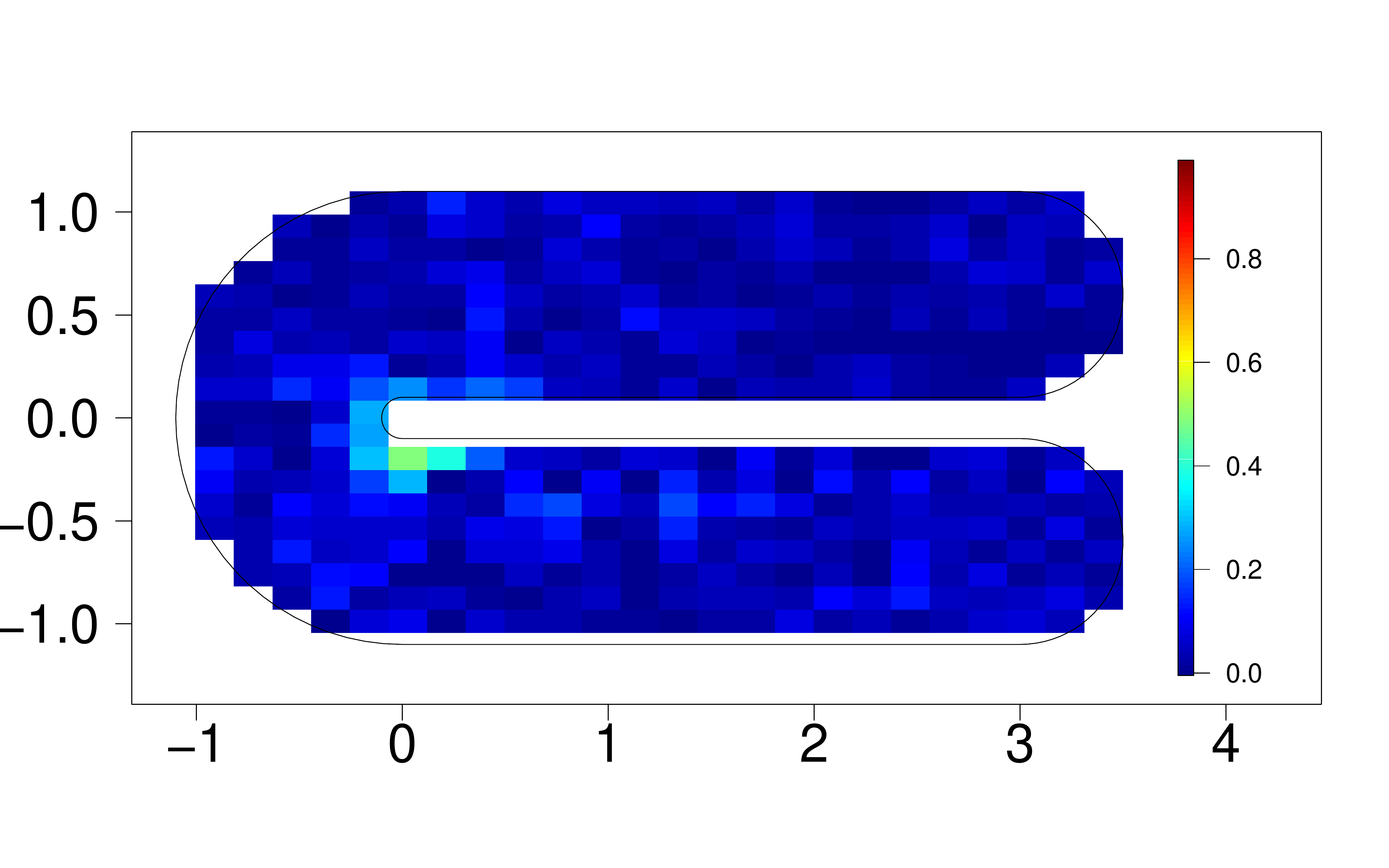}}
	\caption{Prediction error for correlation matrices estimated via RDD-MK method, computed as Riemannian distance between prediction and target, $d_R(\boldchi_{\s_0}, \boldchi^*_{\s_0})$.}\label{fig:cor-errors}
\end{figure}

\begin{figure}
	\centering
	\includegraphics[width=.49\textwidth]{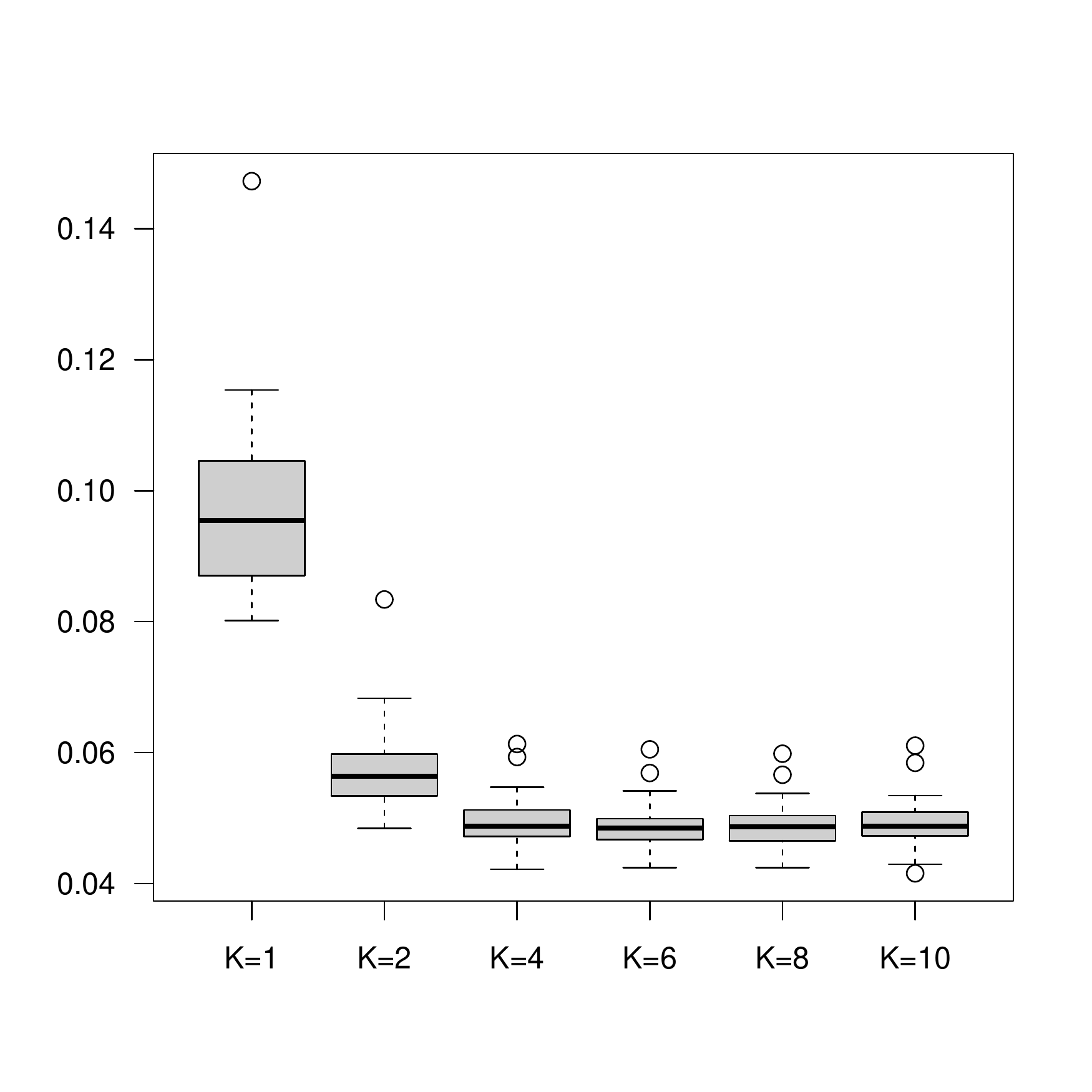}
	\includegraphics[width=.49\textwidth]{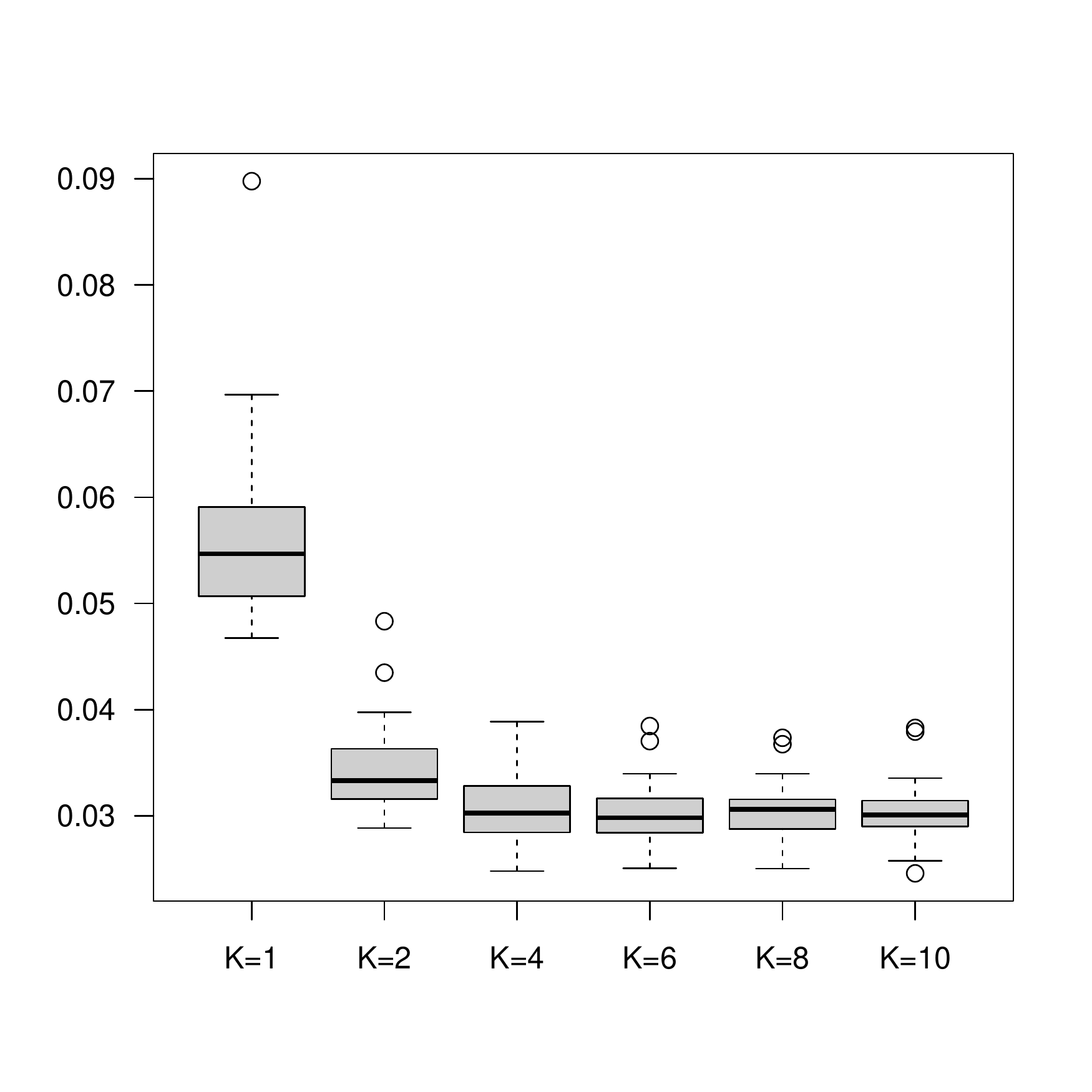}\\
	\caption{Boxplots of prediction errors for the correlation matrices field for different values of the number of tiles used in the random domain decomposition algorithm. Left: prediction errors computed using Cholesky distance. Right: prediction errors computed as square difference between correlation.}\label{fig:sim_cor}
\end{figure}

We here illustrate a second simulation study where the correlation matrices are the objects of interest and we want therefore to use the geometry of the Cholesky manifold in the random domain decomposition algorithm.
Data are simulated on the same C-shaped domain described in Section \ref{subsec:covsim} and using the same generative process to simulate $2\times2$ covariance matrices, which are then transformed into the corresponding correlation matrices. These are then treated as the given data objects. It is worth to notice that this means that even locally the generative process does not coincide with the local tangent space model on the Cholesky manifold that we are going to use in the algorithm, although we expect this to be a useful approximation. We simulated the random field on the grid points $\mathcal{G}$ of the C-shaped domain (see Section \ref{subsec:covsim}) for $30$ times and for each of these replicates we randomly subsampled $100$ grid points to be labelled as observed. We then compute the Cholesky matrix for each of the observed correlation matrix and we apply the RDD-MK algorithm for kriging described in Section \ref{sec:RDD-man}, using the Cholesky manifold geometry discussed in Section \ref{subsubsec:cor} and a varying number of tiles. This results in the kriging prediction of the Cholesky matrices at each grid point, from which the corresponding correlation matrices can be directly obtained. Finally, we compute the prediction error by averaging over all the unobserved locations the square Cholesky distance between the true (simulated) value for the correlation matrix field and the prediction obtained from the random domain decomposition algorithm. Alternatively, since we are in the special case of the $2\times2$ matrices, we also compute the average square difference between true and predicted correlation indices (i.e. the only out of diagonal element of the correlation matrix).  Figure \ref{fig:sim_cor} shows the boxplot of these two types of errors across the simulated replicates for the various number of tiles set in the algorithm. Both metrics suggest the use of $K=4$ tiles for the kriging prediction and this leads to a relevant decrease of the prediction error with respect to the stationary case of \citep{PigoliEtAl2016} ($K=1$). To investigate the reason for this improvement, Figures \ref{fig:results-cor} and \ref{fig:cor-errors} show the results (predicted field and prediction error, respectively) for the first replicate in the simulation study for $K=1,2$ and $4$. It can be again seen how the stationary approach suffers where the domain is far from Euclidean and the global approach incorrectly combines information from parts of the domain that are not close in the correct geometry. On the other hand, the fact that there is a gain from using more than $2$ tiles seems to suggest that a more localised approach is necessary to account for the non Euclidean nature of the data beyond what would be needed because of the complex topology of the domain.

\section{A case study: analysis of associations between aquatic variables in the Chesapeake Bay}\label{sec:case-study}
As an illustrative example, we here consider the spatial prediction of the association between aquatic variables in the Chesapeake Bay. This is the largest estuarine systems in the United States and one of its most productive and complex ecosystem, whose restoration and protection is in charge of the Chesapeake Bay Program (CBP). Amongst the aquatic variables monitored by CBP, the concentration of dissolved oxygen (DO) in water is of particular interest, as it is key to guarantee the life of most marine species. An RDD approach for the spatial analysis and prediction of the probability density functions of DO in the Chesapeake Bay is reported in \citep{MenafoglioEtAl2018}.
It is widely recognized that DO is influenced by temperature: for instance, the solubility of oxygen decreases as the water temperature (WT) increases (negative correlation). We here aim to study the spatial variation within the Bay of the variability of DO and of WT, and of their covariation.

\begin{figure}
  \centering
  \subfigure[Covariance matrices at the sampling sites]{\includegraphics[width=0.45\textwidth]{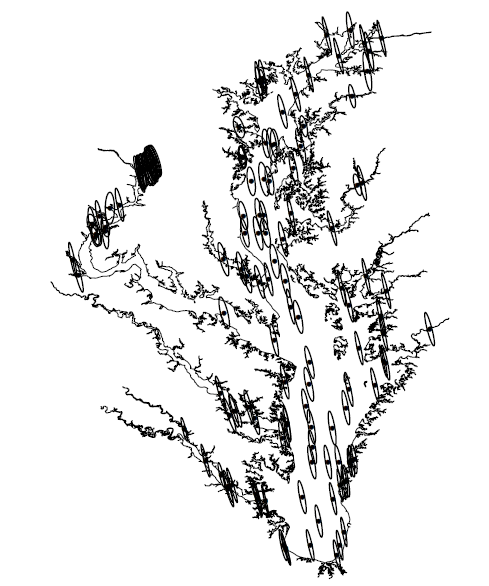}\label{subfig:chesa-data-cov}}
  \subfigure[Domain representation through constrained Delaunay triangulation]{\includegraphics[width=0.45\textwidth]{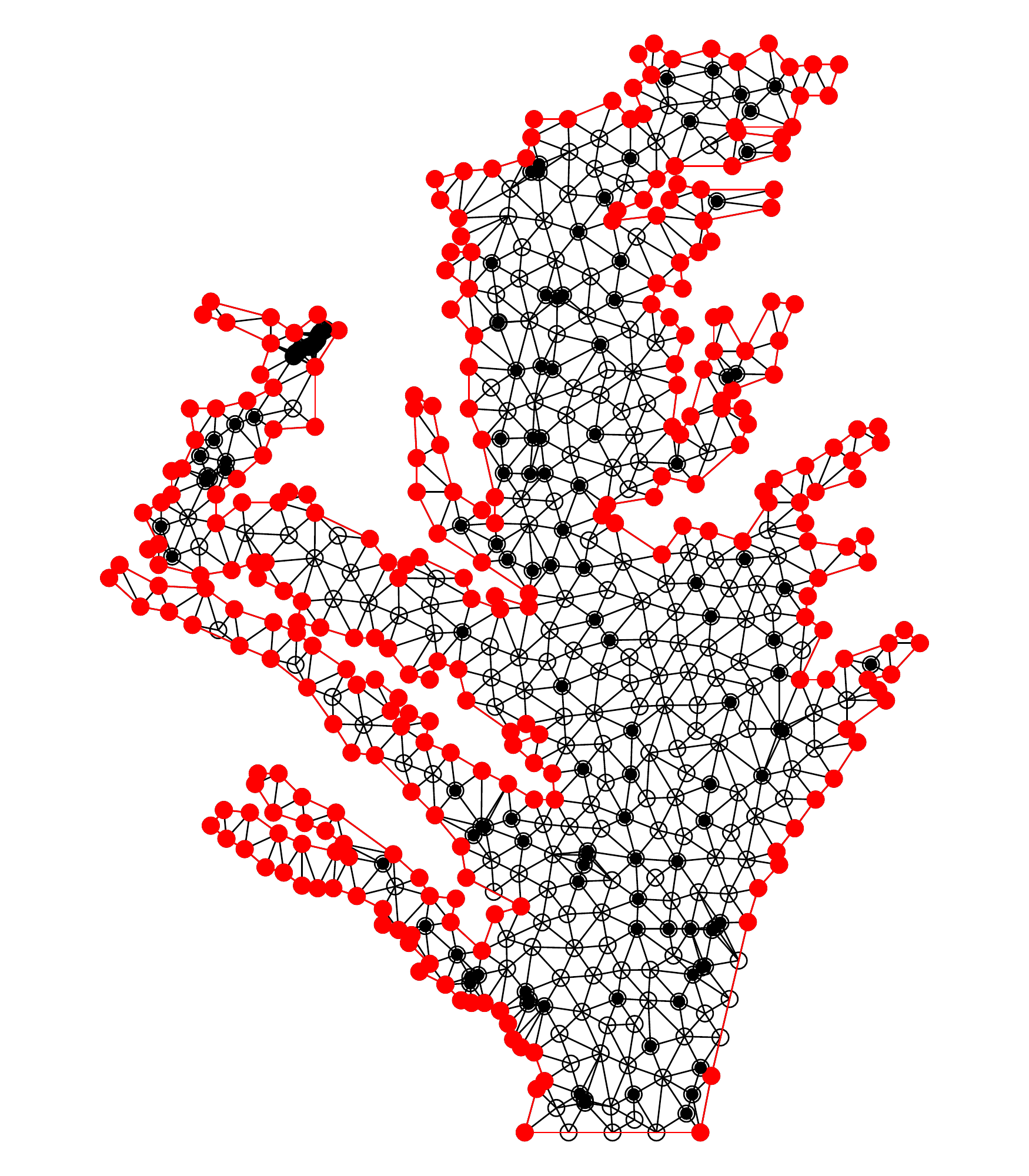}\label{subfig:chesa-data-mesh}}
  \caption{Chesapeake Bay data: covariance matrices between dissolved oxygen and water temperature at the 144 data locations in the Bay. In panel (a) data were represented through the ellipses generated by the associated quadratic forms (radius of ellipses: 0.01). }\label{fig:chesa-data}
\end{figure}

The dataset we consider consists of the sample covariance matrices between DO and WT estimated at 144 locations within the Bay. These were computed on the basis of the joint measurements of (DO, WT) available at those locations along the period 1990-2006 [source: US Environmental Protection Agency Chesapeake Bay Program (US EPA-CBP)]. Although the original database was relatively large, a high number of missing data was present at several locations; to control for the variability of the estimated covariance matrices, only the locations with more than 10 measurements were considered for the analysis. For the purpose of this illustration, we assume the covariance matrices of the random vector (DO, WT) to be constant over time, and we neglect the uncertainty associated with their estimates. An in-depth analysis of these aspects could be the scope of future work. Figure \ref{subfig:chesa-data-cov} reports the covariance matrices estimated at the 144 data locations.

As noted by several authors (e.g., \citet{JensenEtAl2006}, \citet{MenafoglioEtAl2018}), the complex topology of the Bay -- characterized by irregular boundaries and non-convexities -- prevents the use of a global stationary model for the phenomenon based on a Euclidean metric for the spatial domain. Indeed, the distance among sites at which aquatic variables are recorded should be measured in terms of \emph{water distance}, rather than their Euclidean distance. For this reason, we shall consider on the spatial domain the graph-distance based on a mesh representation of the Bay, which provides an approximation of the water distance among locations. We thus employ the simplified spatial domain used in \citet{MenafoglioEtAl2018}, and represent the latter spatial domain through a constrained Delaunay triangulation with vertices at the spatial locations. To define the boundaries and refine the quality of the triangulation, additional vertices were added (red and empty symbols in Figure \ref{subfig:chesa-data-mesh}).

\begin{figure}
  \centering
  \subfigure[$K=1$]{\includegraphics[width=0.3\textwidth]{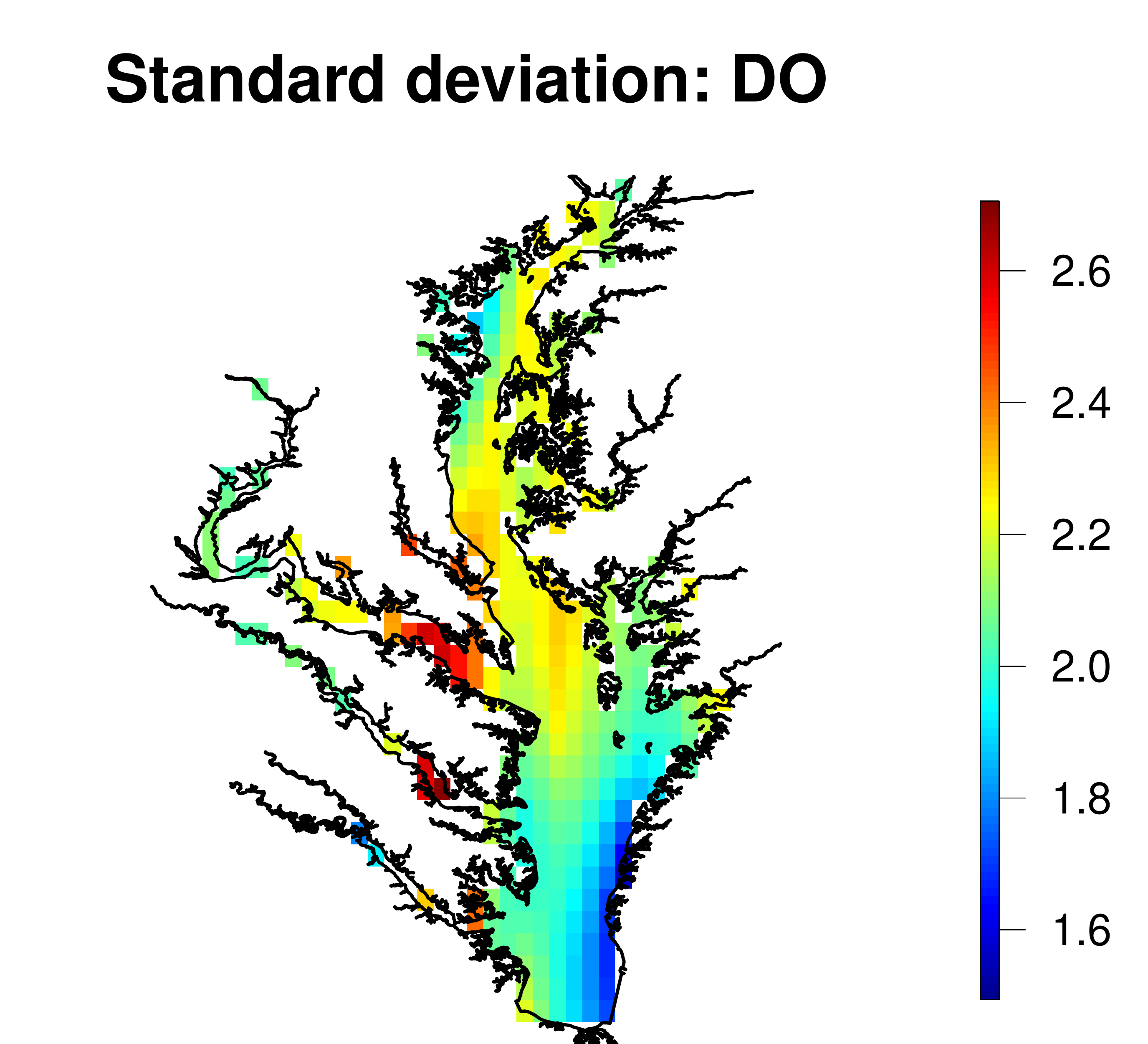}\includegraphics[width=0.3\textwidth]{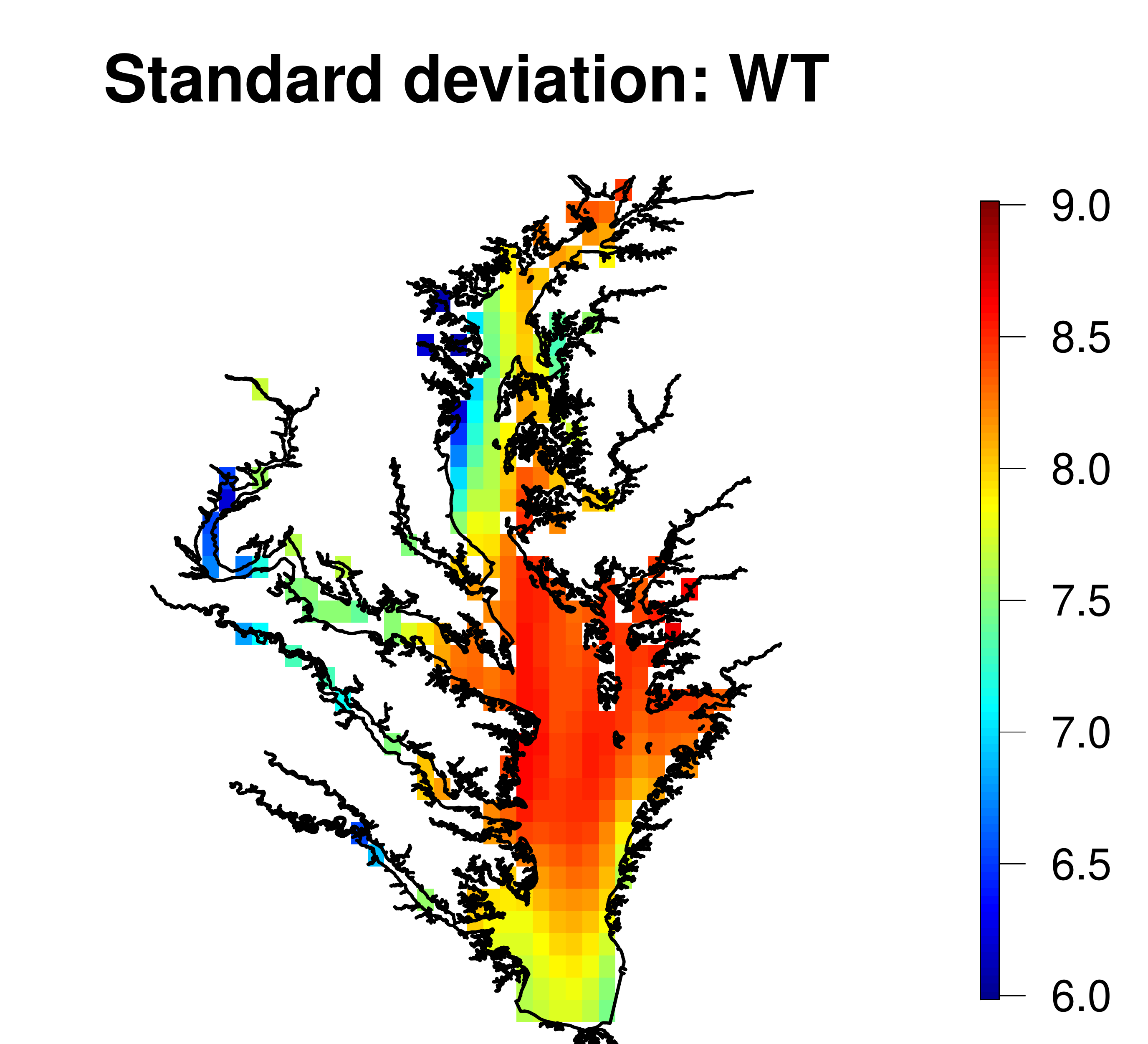}
  \includegraphics[width=0.3\textwidth]{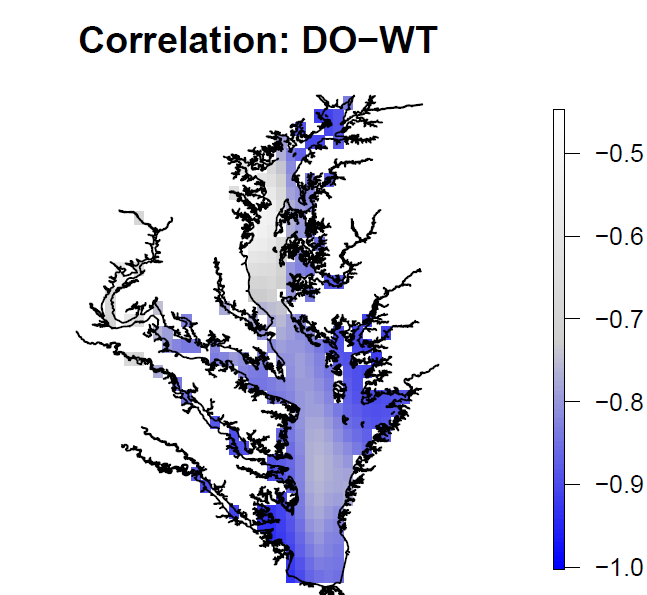}}
  \subfigure[$K=6$]{\includegraphics[width=0.3\textwidth]{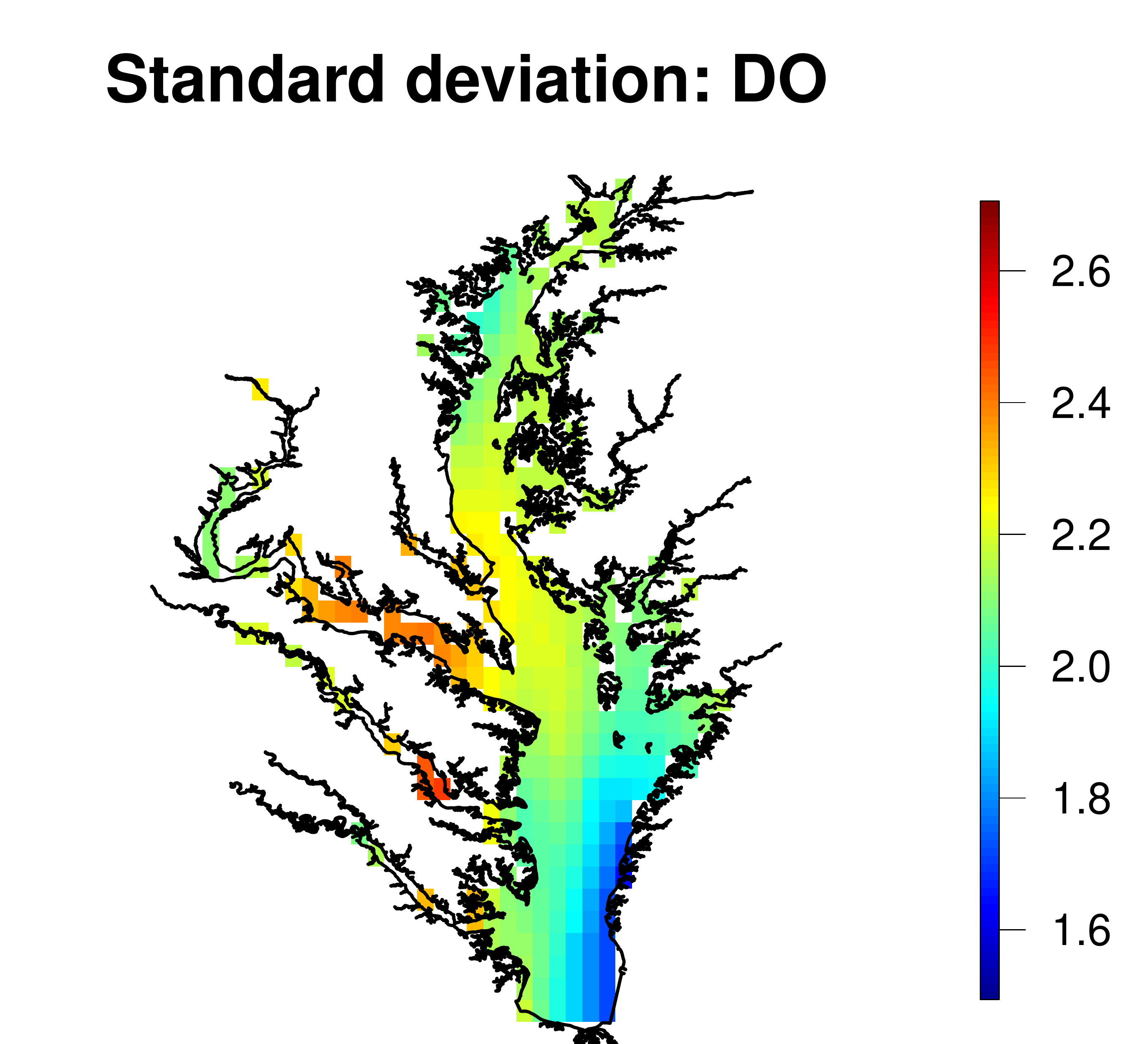}\includegraphics[width=0.3\textwidth]{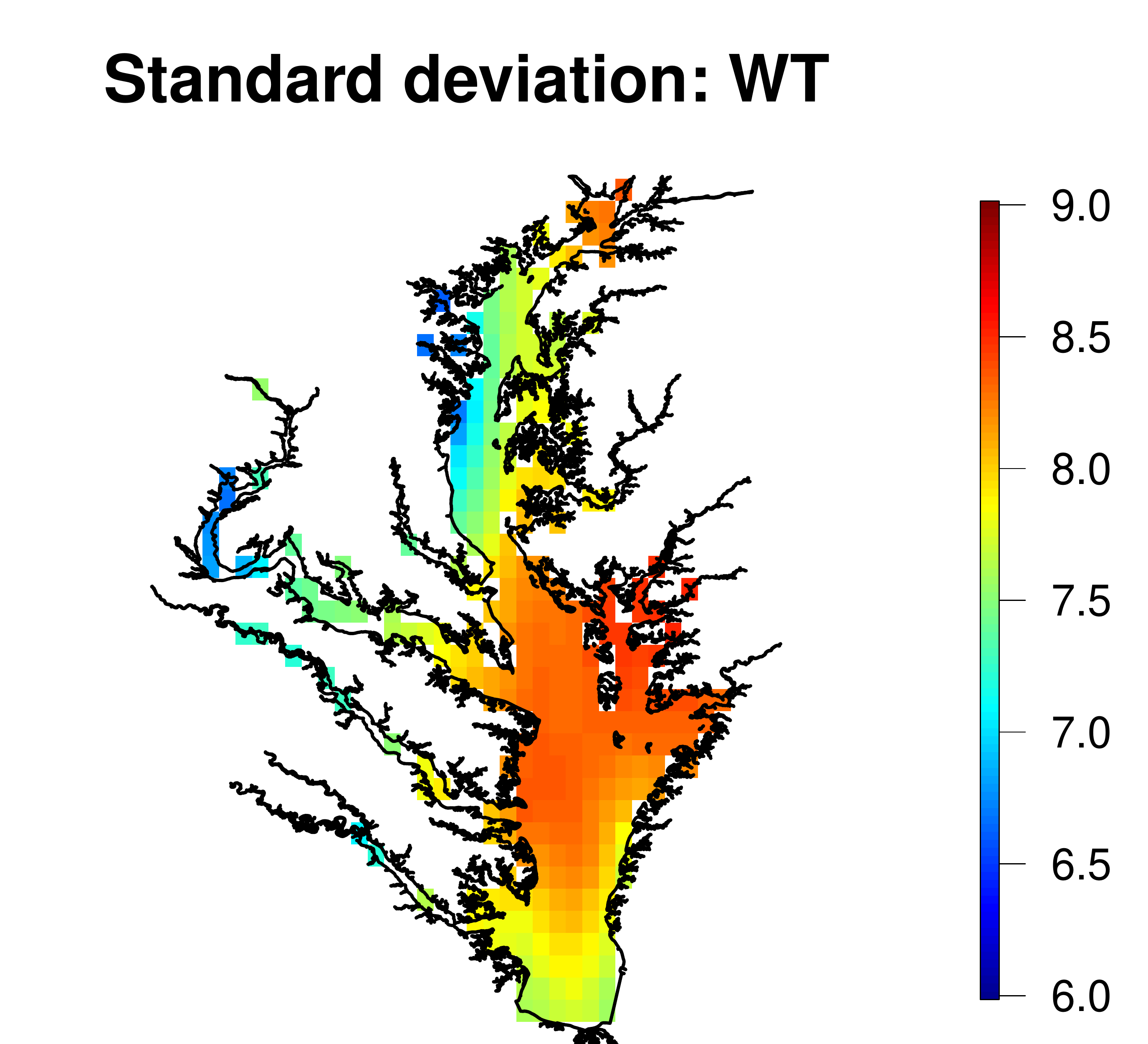}
  \includegraphics[width=0.3\textwidth]{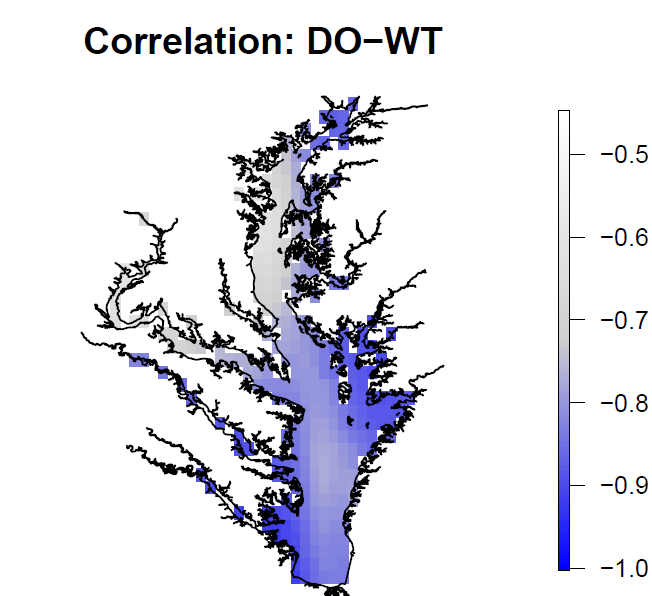}}
  \subfigure[$K=10$]{\includegraphics[width=0.3\textwidth]{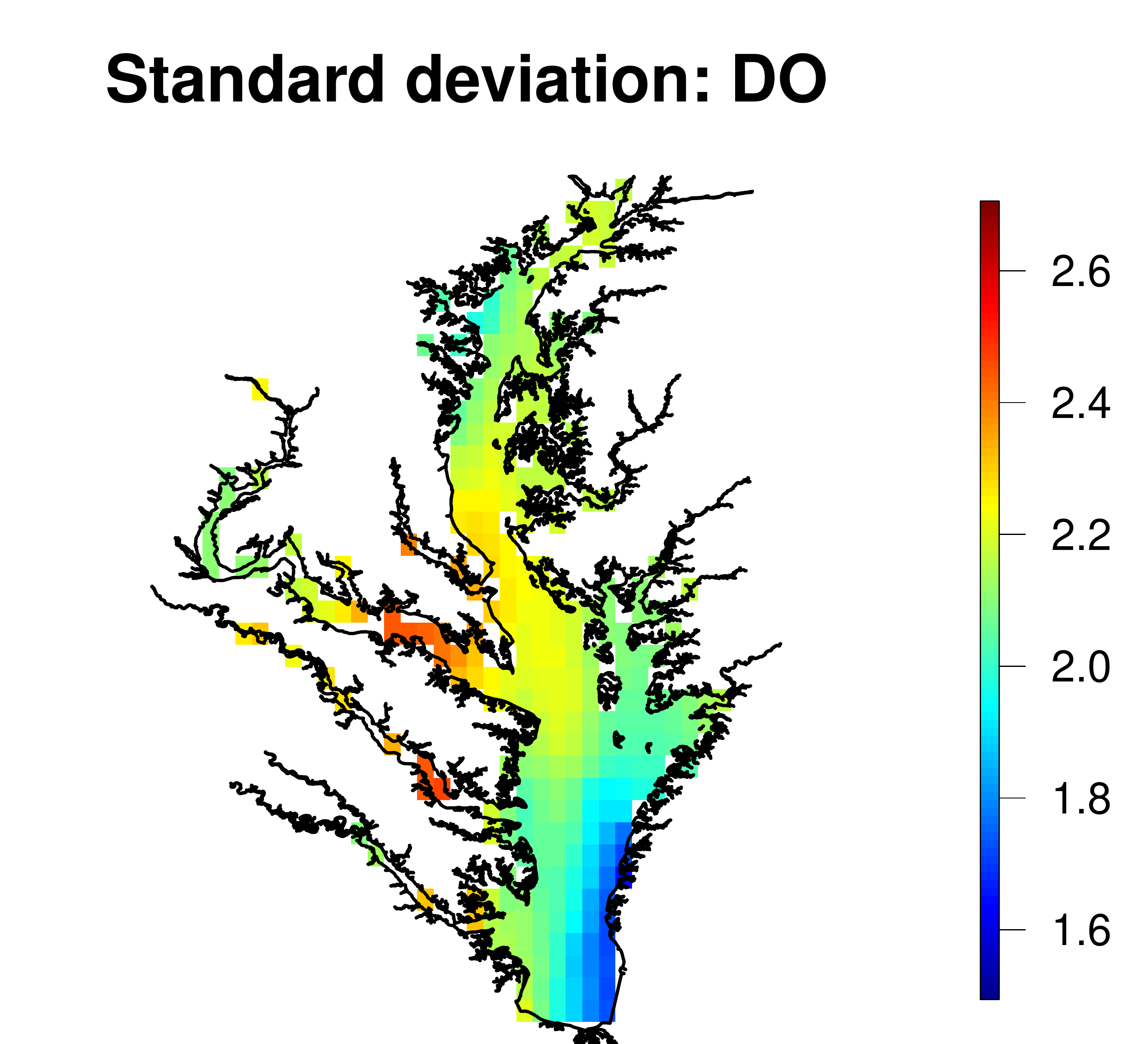}\includegraphics[width=0.3\textwidth]{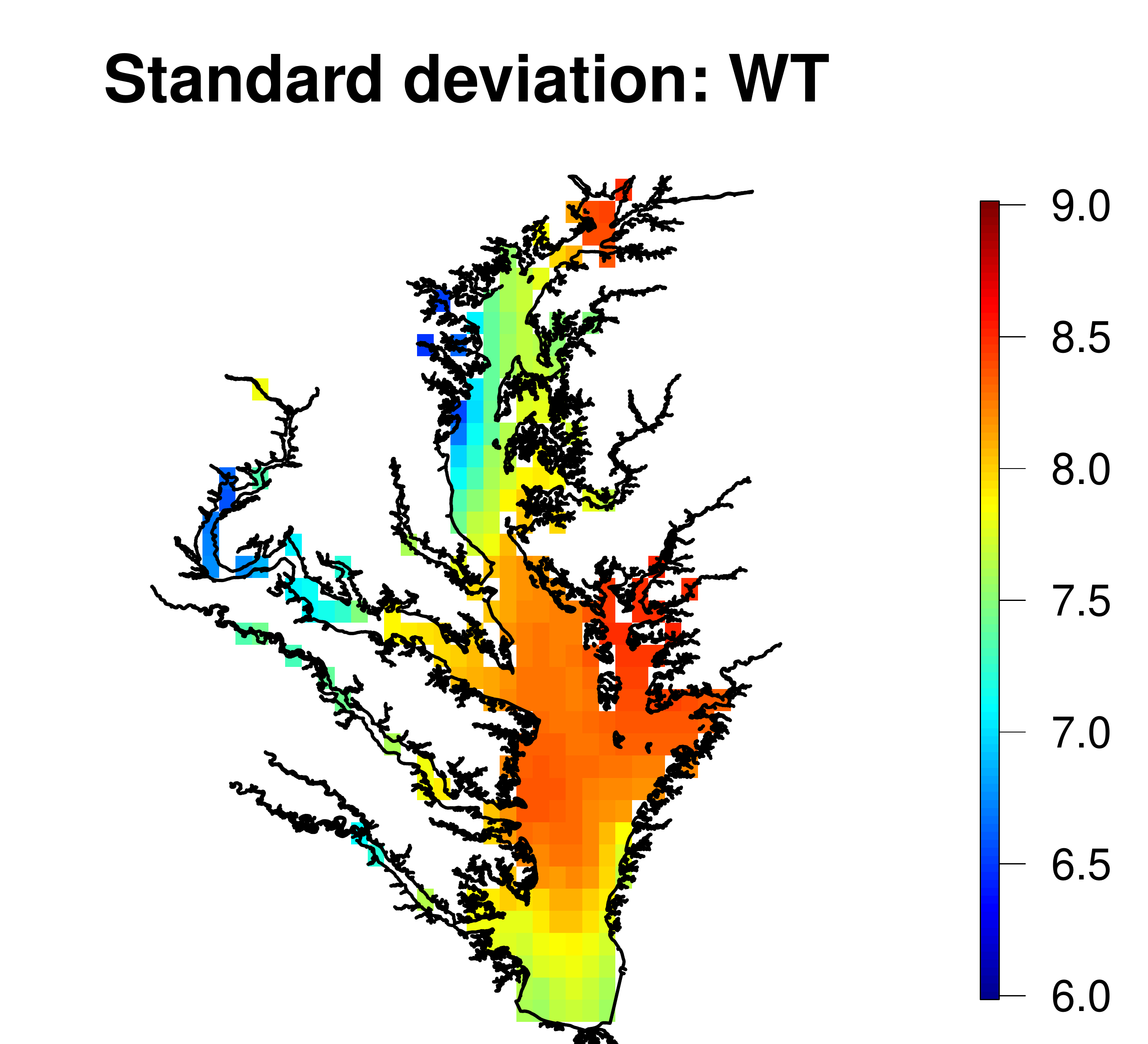}
  \includegraphics[width=0.3\textwidth]{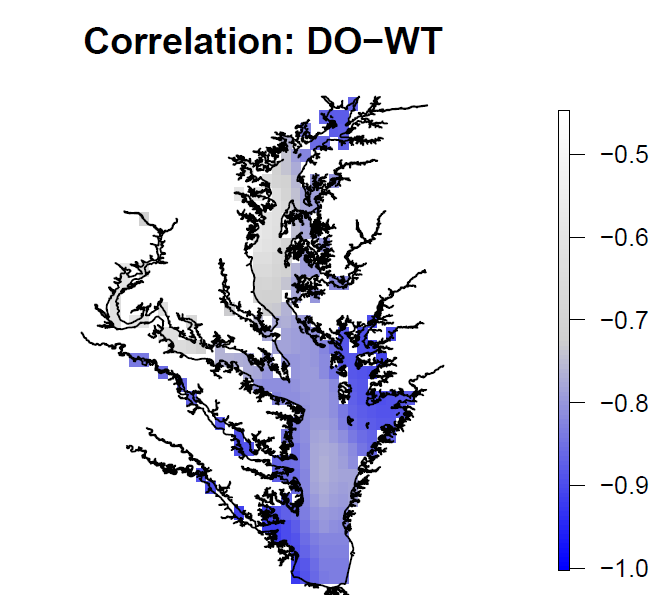}}
  \caption{Prediction of covariance matrices between DO and WT in the Chesapeake Bay.}\label{fig:chesa-results-cov}
\end{figure}

We thus applied the RDD-MK method to the covariance matrices, by using the method detailed in Subsection \ref{subsubsec:cov} with a Gaussian kernel with bandwidth $\epsilon = 100$ for the variogram estimation. Figure \ref{fig:chesa-results-cov} reports the results obtained for $K=1, 6, 10$, as aggregation of $B=100$ bootstrap iterations. Results are reported through the maps of the standard deviations of DO and WT, and the map of the correlation between the two variables. Graphical inspection of Figure \ref{fig:chesa-results-cov} suggests that the standard deviations predicted via RDD-MK with $K=6,10$ tend to be smaller than those obtained with $K=1$, for both variables DO and WT; these differences appear both in the main branch and in the lateral ones. Regarding the correlations -- which are all negative, as expected -- those predicted for $K=6,10$ are generally smaller, in absolute value, than those obtained with $K=1$. These differences appear especially in the main left branch, where predictions for $K=1$ seem to result from an inappropriate smoothing across the upper and central left branches of the Bay. Note particularly that, by increasing $K$, the results appear to better reflect the topology of the domain. Leave-one-out cross-validation results, reported in Table \ref{tab:CV-cov}, suggests a slight improvement of predictions for $K=8,10$ with respect to $K=1$. There is therefore some indication in favour of using a relatively large number of tiles, although we should take these results with some care, since cross-validation is not completely reliable in this context, because of the spatial dependence among data and the possible presence of influential data in the sampling design. 
We also remark that the choice of $K$ is informed by the number of available observations. Indeed, by increasing $K$, the average size of the tiles is decreased. For the case study considered here -- which is based on a relatively small number of sites -- using a number of tiles higher than $K=10$ is likely to be associated with a strong uncertainty in the results. The three cases displayed in Figure \ref{fig:chesa-results-cov} are thus representative of cases of a low, moderate and high number of tiles.

\begin{table}
  \centering

\begin{tabular}{ccccccc}
  \toprule
  $K$   & 1 & 2 & 4 & 6 & 8 & 10 \\\midrule
  Mean  & 0.501 & 0.527 & 0.519 & 0.508 & 0.497 & \textbf{0.486} \\
  Median & 0.425& 0.471 & 0.464 & 0.443 & 0.430 & \textbf{0.414} \\
  \bottomrule
\end{tabular}
   \caption{Cross-validation results on the Chesapeake Bay case study, when the data objects are covariance matrices. The entries are the mean and the median of the square prediction error, assessed by leave-one-out cross-validation, for $K\in\{1,2,4,6,8,10\}$. The minimum over $K$ is highlighted in bold.}\label{tab:CV-cov}
\end{table}

\begin{figure}
  \centering
  \subfigure[$K=6$]{\includegraphics[width=0.4\textwidth]{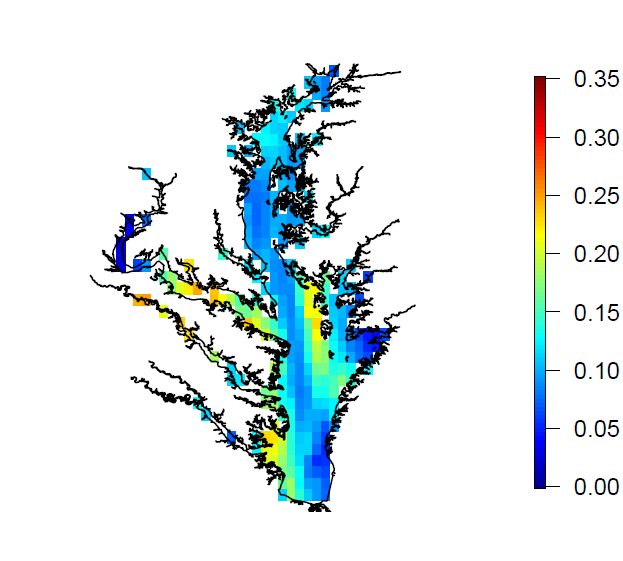}\label{subfig:chesa-bootvar-cov6}}\hspace*{5mm}
  \subfigure[$K=10$]{\includegraphics[width=0.4\textwidth]{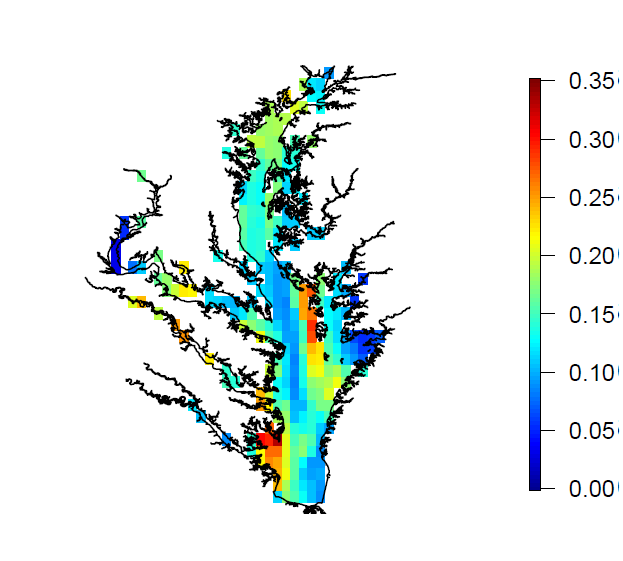}\label{subfig:chesa-bootvar-cov10}}\hspace*{5mm}
  \caption{Bootstrap variance of RDD-MK for covariance matrices between DO and WT in the Chesapeake Bay.\label{fig:chesa-bootvar-cov}}
\end{figure}

Figure \ref{fig:chesa-bootvar-cov} reports the bootstrap variance associated with the predictions in Figure \ref{fig:chesa-results-cov}, for $K=6,10$. The latter was computed as
\[
\varsigma^2(\s_0) = \frac{1}{B}\sum_{b=1}^B d^2(\boldchi_{\s_0}^{*,b}, \boldchi_{\s_0}^{*}),\]
where $\boldchi_{\s_0}^{*,b}$ is the kriging prediction at location $\s_0$ at iteration $b$, for $\s_0$ in $D$ and $b=1,...,B$, whereas $\boldchi_{\s_0}^{*}$ is the aggregated predictor. The bootstrap variance provides indications on the areas of higher instability of the predictor along the bootstrap replicates (for further discussion see \citet{MenafoglioEtAl2018}). Clearly, for $K=1$, the bootstrap variance is uniformly zero over the field. Inspection of Figure \ref{fig:chesa-bootvar-cov} suggests that the regions of higher variability are located along the areas of the main branch of the Bay, characterized by particularly irregular boundaries, or in correspondence of isolated data within the left branches. These areas may be associated with a more complex spatial structure or influent data as in the left branches; similar results were also obtained by \citep{MenafoglioEtAl2018}.
\bigskip

\begin{figure}
	\centering
	\subfigure[$K=1$]{\includegraphics[width=.3\textwidth]{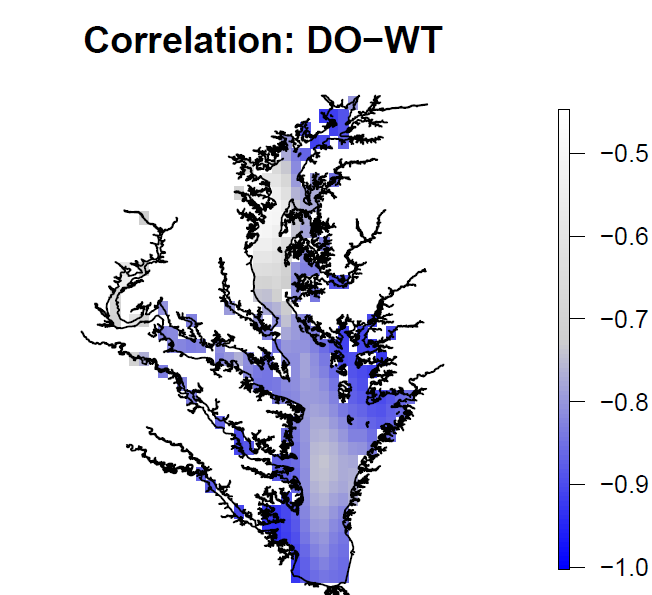} 
    }
	\subfigure[$K=6$]{\includegraphics[width=.3\textwidth]{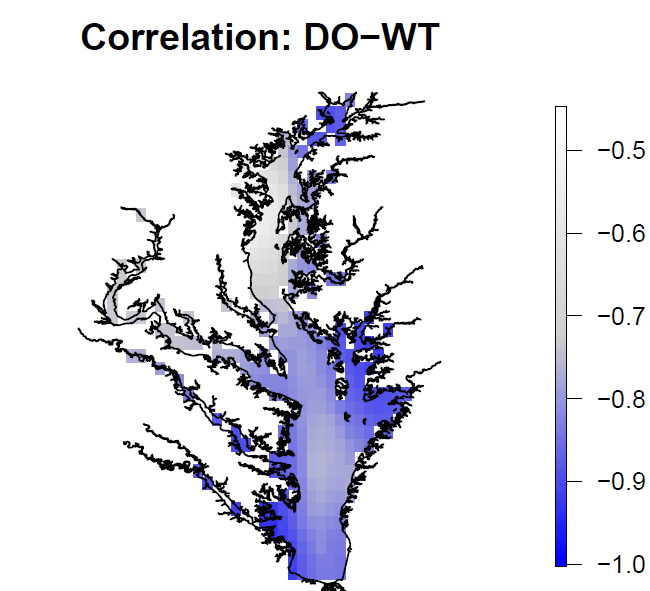} 
}
	\subfigure[$K=10$]{\includegraphics[width=.3\textwidth]{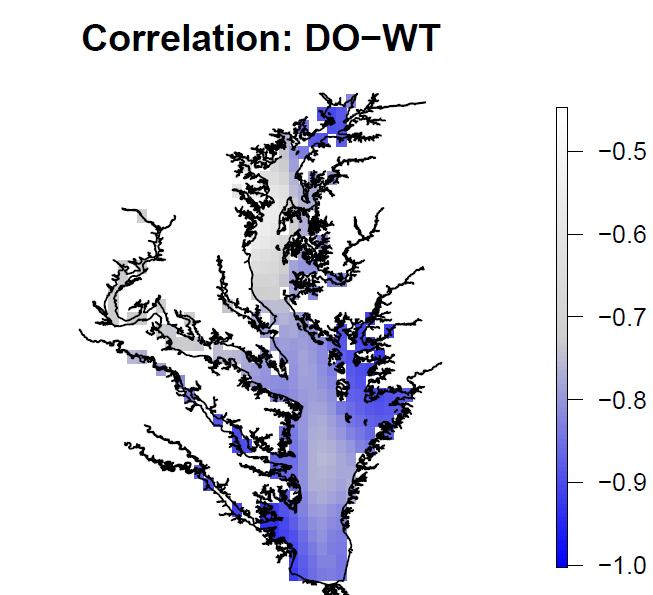} 
}
	\caption{Spatial prediction of the correlation between DO and WT via RDD-MK applied on the correlation matrices. 
}\label{fig:chesa-results-cor}
\end{figure}

For the sake of comparison, Figure \ref{fig:chesa-results-cor} displays the correlation indices predicted via RDD-MK when applied directly to the correlation matrices, by following the method of Subsection \ref{subsubsec:cor}. Results are reported for $K=1,6,10$ and with the same parameter settings as before. Comparison of Figures \ref{fig:chesa-results-cov} and \ref{fig:chesa-results-cor} suggests that the results of the two analyses are indeed consistent.

\section{Conclusion}\label{sec:concl}
This is a paper in Object Oriented Spatial Statistics (O2S2), where we presented an operational strategy for the prediction of manifold data observed on a spatial domain, when spatial dependence is an important issue of the analysis.

In a previous work \citep{PigoliEtAl2016}, we developed a Kriging approach for manifold data, suitable for situations in which the data variability is limited and the random spatial field generating the data is stationary.  Expanding to a more general line of attack able to deal with a non-stationary random field is highly non trivial. To meet this challenge, we have here proposed a scheme that localizes the analysis by generating an ensemble of local models. The strategy is founded on the RDD approach -- originally introduced in \citet{MenafoglioEtAl2018} -- which repeatedly splits the spatial domain of interest according to independent realizations of a random partition. At each iteration of the algorithm, a local analysis is conducted in each element of the partition, where the variability of the observed data can be reasonably assumed to be limited and the field generating them to be stationary. These local analyses are finally aggregated into a global one.

The advantages of the RDD-MK approach for O2S2 are twofold. First, the possibility to deal with data generated by random fields which are only locally stationary. Second, the potential for the analysis of data observed on complex spatial domains, non Euclidean at a global scale, but locally approximated by simple Euclidean subdomains. These include spatial domains with holes, barriers and more general non-convexities where the appropriate notion of closeness is not captured by the Euclidean distance. The Chesapeake Bay estuarine system here considered (Section \ref{sec:case-study}) is a real world paradigmatic example. At this point in the discussion, it is worth recalling that kriging data observed on non-Euclidean domains is still an open challenge in geostatistics, due to the absence of valid variogram models.

We have applied our operational strategy to the analysis and prediction or random fields of positive definite symmetric matrices, namely covariance or correlation matrices, observed on complex spatial domains. Covariances and correlations have been embedded in different Riemannian manifolds; the two in silico case studies, one for covariances and the other for correlations, provide support to a proper Object Oriented perspective on data analysis.

Finally, as in \citet{MenafoglioEtAl2018}, we point out that by localizing the analysis through a random domain partition of the spatial domain, we are introducing an exogenous source of variability. In fact, the bagging scheme illustrated in the paper naturally leads to the evaluation of a bootstrap variance estimated in each site of the spatial prediction grid, and this in turns identifies locations of larger instability of the predictor, discontinuities and, more generally, subdomains where the assumption of local stationarity could not be viable. Nonetheless, further research is still needed to decouple this source of variability from the natural endogenous variability of the phenomenon under study. 

\section*{Acknowledgements}
The authors wish to thank Ilaria Sartori and Luca Torriani for supporting them in writing in C++ parts of the code used to fit the local models, thus speeding up considerably our original R code and making possible extensive simulation studies.

\bibliography{biblioRDD}
\end{document}